\documentclass[twocolumn,tighten]{aastex631}
\usepackage{braket,multirow,amsmath,amssymb}
\usepackage{array,multirow}
\usepackage{color,float}
\usepackage{bm}

\begin{document}

\title{Gravitational Microlensing Rates in Milky Way Globular Clusters}

\author[0000-0003-4412-2176]{Fulya K{\i}ro\u{g}lu}
\affil{Department of Physics \& Astronomy, Northwestern University, Evanston, IL 60208, USA}
\affil{Center for Interdisciplinary Exploration \& Research in Astrophysics (CIERA), Northwestern University, Evanston, IL 60208, USA}
\correspondingauthor{Fulya K{\i}ro\u{g}lu}
\email{fulyakiroglu2024@u.northwestern.edu}

\author[0000-0002-9660-9085]{Newlin C. Weatherford}
\affil{Department of Physics \& Astronomy, Northwestern University, Evanston, IL 60208, USA}
\affil{Center for Interdisciplinary Exploration \& Research in Astrophysics (CIERA), Northwestern University, Evanston, IL 60208, USA}

\author[0000-0002-4086-3180]{Kyle Kremer}
\affil{TAPIR, California Institute of Technology, Pasadena, CA 91125, USA}
\affil{The Observatories of the Carnegie Institution for Science, Pasadena, CA 91101, USA}

\author[0000-0001-9582-881X]{Claire S. Ye}
\affil{Department of Physics \& Astronomy, Northwestern University, Evanston, IL 60208, USA}
\affil{Center for Interdisciplinary Exploration \& Research in Astrophysics (CIERA), Northwestern University, Evanston, IL 60208, USA}

\author[0000-0002-7330-027X]{Giacomo Fragione}
\affil{Department of Physics \& Astronomy, Northwestern University, Evanston, IL 60208, USA}
\affil{Center for Interdisciplinary Exploration \& Research in Astrophysics (CIERA), Northwestern University, Evanston, IL 60208, USA}

\author[0000-0002-7132-418X]{Frederic A. Rasio}
\affil{Department of Physics \& Astronomy, Northwestern University, Evanston, IL 60208, USA}
\affil{Center for Interdisciplinary Exploration \& Research in Astrophysics (CIERA), Northwestern University, Evanston, IL 60208, USA}

\begin{abstract}
Many recent observational and theoretical studies suggest that globular clusters (GCs) host compact object populations large enough to play dominant roles in their overall dynamical evolution. Yet direct detection, particularly of black holes and neutron stars, remains rare and limited to special cases, such as when these objects reside in close binaries with bright companions. Here we examine the potential of microlensing detections to further constrain these dark populations. Based on state-of-the-art GC models from the \texttt{CMC Cluster Catalog}, we estimate the microlensing event rates for black holes, neutron stars, white dwarfs, and, for comparison, also for M~dwarfs in Milky Way GCs, as well as the effects of different initial conditions on these rates. Among compact objects, we find that white dwarfs dominate the microlensing rates, simply because they largely dominate by numbers. We show that microlensing detections are in general more likely in GCs with higher initial densities, especially in clusters that undergo core collapse. We also estimate microlensing rates in the specific cases of M22 and 47~Tuc using our best-fitting models for these GCs. Because their positions on the sky lie near the rich stellar backgrounds of the Galactic bulge and the Small Magellanic Cloud, respectively, these clusters are among the Galactic GCs best-suited for dedicated microlensing surveys. The upcoming 10-year survey with the Rubin Observatory may be ideal for detecting lensing events in GCs.
\end{abstract}

\section{Introduction} \label{sec:intro}

\indent Constraining the demographics of compact object populations in globular clusters (GCs) has been of high interest in astronomy for several decades. Compact objects---stellar-mass black holes (BHs), neutron stars (NSs), and white dwarfs (WDs)---are the key ingredients for a wide array of sources and phenomena observed in clusters, including X-ray binaries \citep[][]{Katz_1975,Clark_1975,Verbunt_1984,Heinke_2005,Ivanova_2013,Giesler_2018,Kremer_2018a}, millisecond radio pulsars \citep[][]{Lyne_1987,Sigurdsson_1995,Camilo_2005,Ransom_2008,Chomiuk_2013,Shishkovsky_2018,Fragione_2018c,Ye_2019}, fast radio bursts \citep[][]{Kirsten_2021,Kremer_2021_FRB}, and gravitational wave events \citep[][]{Moody_2008,Banerjee_2010,Bae_2014,Ziosi_2014,Rodriguez_2015,Rodriguez_2016,Askar_2016,Banerjee_2017,Hong_2018,Fragione_2018,Samsing_2018,Rodriguez_2018,Zevin_2019,Kremer_2019d}. Populations of compact objects also greatly impact the overall dynamical evolution of GCs; in particular, stellar BHs can quickly concentrate in the cluster core due to dynamical friction and subsequently heat the cluster through frequent binary-mediated encounters \citep{Spitzer_1969,Heggie_2003,Breen_2013,Kremer_2019b}. Furthermore, massive WDs can dominate the central regions of \textit{core-collapsed} GCs, also driving their dynamical evolution \citep[e.g.,][]{Kremer_2020,Rui_2021b,Kremer_2021}.

Observations and theory alike strongly indicate the existence of compact object populations in GCs. Milky Way GCs contain several known stellar-mass BH binary candidates detected via radio and X-ray observations, including in NGC~4472 \citep[]{Maccarone_2007}, M22 \citep{Strader_2012}, M62 \citep{Chomiuk_2013}, 47~Tuc \citep{Miller-Jones_2015}, and M10 \citep{Shishkovsky_2018}, as well as those detected via radial velocity measurements in NGC~3201 \citep{Giesers_2018,Giesers_2019}. 
Numerical simulations of GCs reinforce this observational evidence by demonstrating that Milky Way GCs can retain large populations of stellar-mass BHs up to the present day \citep[e.g.,][]{Merritt_2004,Mackey_2007,Mackey_2008,Breen_2013,Morscher_2015,Peuten_2016,Chatterjee_2017a,Chatterjee_2017b,Askar_2018,Kremer_2018b,Kremer_2020,Weatherford_2018,Weatherford_2020,Antonini_2020}.
Meanwhile, observations of millisecond pulsars and low-mass X-ray binaries with neutron star accretors suggest that Milky Way GCs may on average contain hundreds of NSs each \citep[e.g.,][]{Ivanova_2008,Kuranov_2006}. Observations of WDs in GCs are also important; notably they enable stronger predictions of their host clusters' ages \citep[][]{Hansen_2002,Hansen_2007,Hansen_2013} and distances \citep{Renzini_1996}. WDs are also observable as cataclysmic variables via their variability, emission lines, colours and X-ray emission \citep[i.e.,][]{Knigge_2012}. However, with their relatively large distances, GC cataclysmic variables are 10-100 times less bright than nearby ones observed in the Galactic field.

Yet the aforementioned observations are largely limited to special cases, particularly for BHs. Although NSs are detectable as pulsars and the younger, luminous end of the WD sequence is observable in some nearby clusters \citep[e.g.,][]{Richer_1995,Richer_1997,Cool_1996}, compact objects in GCs have otherwise only been directly detected in binaries via their influence on a luminous companion. This can be problematic when trying to use existing observations to constrain bulk properties of compact object populations in GCs. For example, dynamical simulations of GCs establish that the number of cluster BHs residing in binaries with luminous companions does not correlate with the total number of BHs in a cluster \citep{Chatterjee_2017b,Kremer_2018a}. The apparent BH mass distribution in clusters---useful for constraining supernova and collision physics as well as the cluster initial mass function and star formation history---is susceptible to bias if based solely on observations from binaries. In particular, mass segregation tends to cause the most massive BHs to form binaries with other BHs, not with observable stellar companions; the inferred BH mass distribution from observable BH binaries with a stellar companion could therefore be biased toward lower masses. Thus, in order to better constrain properties of the complete population of compact objects in clusters, additional observational strategies are necessary. Compact object detection through gravitational microlensing may serve such a purpose.

Because the fine alignment required to produce a microlensing event is rare and short-lived, searches for these events are challenging; for instance, early searches by the Massive Astrophysical Compact Halo Object (MACHO) collaboration revealed only three microlensing events by Galactic halo objects despite monitoring nearly $10^7$ stars in the Large Magellanic Cloud for over a year \citep[][]{Alcock_1995,Alcock_1996}.
Early efforts sought to constrain dark compact object populations in the mass range $10^{-7} < M/M_{\odot} < 10^{-1}$ in the Galactic halo and bulge \cite[e.g.,][]{Paczynski_1986,Paczynski_1991,Griest_1991,deRujula_1991}. Since then, however, many microlensing events towards the Galactic bulge have been detected by several collaborations like the Optical Gravitational Lensing Experiment \citep[OGLE;][]{Udalski_1994,Udalski_2003,Udalski_2015}, MACHO \citep[][]{Alcock_2000}, and Microlensing Observations in Astrophysics \citep[MOA;][]{Bond_2001}. Microlensing detections are accelerating as large surveys proliferate. In particular, the OGLE-IV survey now detects around 2000 photometric microlensing events towards the Galactic bulge every year \citep[][]{Udalski_2015}. Most recently, \cite{Sahu_2022} reported the detection of the BH lensing event MOA-11-191/OGLE-11-0462 towards the Galactic Bulge with the BH inferred mass $\sim7.1~M_{\odot}$. The same event has been analyzed in another recent work by \cite{Lam_2022} although they come to a slightly different conclusion than \cite{Sahu_2022} regarding the nature of the object. Thanks to such surveys, the future for microlensing is looking brighter.

The increasing frequency of microlensing detections has led to several recent theoretical studies of microlensing rates for compact objects, including stellar-mass BHs \citep{Lu_2016,Zaris_2020}, intermediate-mass BHs \citep{Kains_2016,Safonova_2007,Kains_2018}, NSs \citep{Dai_2018}, and WDs \citep[][]{McGill_2018}. In particular, \citet{Harding_2017} estimate the microlensing event rate by nearby WD populations to be 30-50 per decade. Due to a small sample of BHs and NSs with well-known distances and proper motions, however, their estimates of the BH and NS rates are less certain.

In general, GCs have well-known distances and velocities, enabling more precise estimates of microlensing rates in these environments. GCs also feature dense populations of compact objects, making their optical depths much higher than in the field of the Milky Way. \cite{Paczynski_1994} originally  proposed microlensing searches on GCs set against the backdrop of the dense Galactic bulge or the Small Magellanic Cloud. The first microlensing event in a GC was observed in M22 and the source star was located in the Galactic bulge \citep{Pietrukowicz_2012}. In addition to lensing of distant background stars, GCs can also produce observable microlensing events of the cluster stars themselves (so-called ``self-lensing''). In this context, using the bright stars in NGC 5139 as sources, \cite{Zaris_2020} estimated the self-lensing rate of BHs in NGC 5139 to be in the range $0.1$--$1 \, \rm{yr}^{-1}$ from their numerical simulations. While a reasonable number of self-lensing events in the dense regions of GCs can in principle occur for a large number of lenses ($\gtrsim 10^4$) and high velocity dispersion, detection requires high resolution imaging of the cluster background stars with powerful telescopes. Moreover, with fewer bright stars contained in GCs compared to the entire field of the Milky Way, many GCs would likely need to be observed to detect some microlensing events \citep{Jetzer_1998}.
  
In this paper, we analyse potential microlensing events in GCs using our realistic \texttt{CMC Cluster Catalog} models \citep{Kremer_2020} with state-of-the-art prescriptions for the formation and kinematics of compact objects. In addition to predicting microlensing rates for compact objects, we also estimate the rates for M~dwarfs, which usually dominate the cluster mass function in our models. In the rate analysis, we give additional attention to the clusters 47~Tuc \citep[][]{Ye_2021} and M22 \citep[][]{Kremer_2019a} given the higher potential for microlensing events provided by these clusters' densely-populated backgrounds on the sky, i.e., the Small Magellanic Cloud (SMC) and the Galactic bulge, respectively.

This paper is structured as follows. In Section~\ref{sec:models}, we describe the computational method used to construct the \texttt{CMC Cluster Catalog} models, as well as the criteria we use to determine which additional \texttt{CMC} models most closely match 47~Tuc and M22. In Section~\ref{sec:rates}, we review the basics of microlensing and explain how we estimate microlensing rates both numerically and analytically from our models. In Section~\ref{sec:results}, we present the microlensing rates due to sources both within the cluster (self-lensing) and outside the cluster (the distant background). Finally, in Section~\ref{sec:discussion}, we summarize and discuss our results.

\section{Modeling dense star clusters} \label{sec:models}

\subsection{CMC Catalog Models}

In this paper, we predict gravitational lensing rates in the Milky Way GCs based on a large grid of cluster simulations recently published as the \texttt{CMC Cluster Catalog} \citep{Kremer_2020}. In particular, we analyse the microlensing event rates of 10 of the catalog's 148 cluster simulations generated with our publicly released Cluster Monte Carlo (\texttt{CMC}) code \cite[][and references therein]{Rodriguez2021}. \texttt{CMC} is a H\'enon-type Monte Carlo code \citep{henon1971monte, henon1971montecluster} that has been continuously developed and improved for over two decades, beginning with \cite{Joshi_2000,Joshi_2001} and \cite{Fregeau_2003}. The Monte Carlo H\'enon method assumes spherical symmetry and orbit averaging, and is parallelized \citep{Pattabiraman_2013}, allowing us to model $10^6$ stars over a Hubble time in a couple days. \texttt{CMC} incorporates all relevant physical processes, such as two-body relaxation, three- and four-body strong encounters \cite[][]{Fregeau_2007}, three-body binary formation \citep[][]{Morscher_2015}, and physical collisions and relativistic dynamics \citep[][]{Rodriguez_2018}. It also includes stellar and binary evolution \cite[][]{Hurley_2000,Hurley_2002} integrated with the \texttt{COSMIC} package for binary population synthesis \cite[][]{Breivik_2020}. We direct the reader to \cite{Rodriguez2021} for a detailed description of \texttt{CMC}'s implementation of all these processes, including up-to-date prescriptions for compact object formation.

All the catalog models adopt the standard \cite{Kroupa_2001} initial mass function (IMF) in the mass range 0.08-150$\,M_{\odot}$. The initial stellar velocities and positions of all objects draw from a King profile with initial central concentration $W_0=5$ \cite[][]{King_1962}. Our models do not include primordial mass segregation, so the initial velocities and positions of all objects do not depend on object mass. In this work, we explore the effect of the initial total number of objects $N$ and the initial cluster virial radius $r_v$ on the microlensing event rates while fixing other initial conditions such as the metallicity $Z=0.1 \,Z_{\odot}$ and the Galactocentric radius $R_{\rm gc}=8$~kpc. Specifically, we use catalog models with initial $N=8\times 10^5$, $1.6\times 10^6$, and $3.2\times 10^6$, while $r_v$ ranges from 0.5 to 4~pc. Because the \texttt{CMC Cluster Catalog} contains only a few models with $N=3.2\times 10^6$, we could only expand our analysis to higher $N$ when using different values of the Galactocentric radius (20~kpc) and metallicity ($Z=0.01 \,Z_{\odot}$). Using a metallicity of $0.01Z_{\odot}$ instead of $0.1Z_{\odot}$ does not significantly impact the microlensing rates as the metallicity does not have a major effect on the number and properties of compact objects below $0.1Z_{\odot}$. It only makes a significant difference as we approach solar metallicity \citep[e.g., see Fig 1 in][]{Kremer_2020}. The primordial binary fraction in each model is $f_b=5\%$; to form binaries, we assign a companion star to a number $N\times f_b$ randomly chosen single stars. The companion masses draw from a flat mass ratio distribution in the range $[0.1, 1]$ \cite[e.g.,][]{Duquennoy_1991}. The initial cluster size is set as the initial virial radius of the cluster, $r_v=GM_{\rm tot}^2/2U$, where $G$ is the gravitational constant, $M_{\rm tot}$ is the total cluster mass, and $U$ is the total cluster potential energy.  As the half-mass relaxation of a cluster depends directly on its virial radius \cite[i.e.,][]{Spitzer_1987},
\begin{equation}
    t_{\rm rh} \sim \frac{M_{\rm tot}^{1/2}}{\braket{m} G^{1/2}\ln{\Lambda}} r_v^{3/2},
\end{equation}
clusters with smaller virial radii evolve faster. Our model clusters truncate at the tidal radius 
\begin{equation}
    r_t = \left(\frac{G M_{\rm tot}}{2 v_{\rm gc}^2}\right)^{1/3} R_{\rm gc}^{2/3},
\end{equation}
where $v_{\rm gc}$ (set to $220\,\rm km\,s^{-1}$) is the circular velocity of the cluster around the Galactic center at Galactocentric distance $R_{\rm gc}$. Tidal stripping of stars due to the Galactic potential follows the \cite{Giersz_2008} energy criterion and is further described by \cite{Chatterjee_2010,Rodriguez2021}.

Table~\ref{table:IC} lists the cluster parameters of all the \texttt{CMC Cluster Catalog} models used in this study, including the theoretical core radius $r_{c,\mathrm{theoretical}}$ \citep[specifically, the density-weighted core radius traditionally used by theorists, e.g.,][]{Casertano_1985} and the half-mass radius $r_h$, which contains half the cluster’s total mass, both at final time $t=12$~Gyr. 

\startlongtable
\begin{deluxetable*}{l|l||cccccc||ccc}
\tabletypesize{\scriptsize}
\tablewidth{0pt}
\tablecaption{Initial cluster parameters for all model GCs\label{table:IC}}
\tablehead{
\colhead{} &
\colhead{Simulation} &
\colhead{$N$}&
\colhead{$r_v$}&
\colhead{$R_{\rm{gc}}$}&
\colhead{$d$}&
\colhead{$Z$}&
\colhead{$f_b$}&
\colhead{$M_{\mathrm{tot},f}$}&
\colhead{$r_{c,f}$}&
\colhead{$r_{h,f}$}\\
\colhead{}&
\colhead{}&
\colhead{$\times10^5$}&
\colhead{pc}&
\colhead{kpc}&
\colhead{kpc}&
\colhead{$Z_{\odot}$}&
\colhead{}&
\colhead{$\times10^5$ $M_\odot$}&
\colhead{pc}&
\colhead{pc}
}
\startdata
1 & \textsc{n8-rv0.5-rg8-z0.1} & 8 & 0.5 & 8 & 8&  0.1 & 0.05& 2.00 & 0.17 & 4.8  \\
2 & \textsc{n16-rv0.5-rg8-z0.1} & 16 & 0.5 & 8 &8& 0.1 &0.05& 4.40 & 0.39 & 4.1  \\
\hline
3 & \textsc{n8-rv1-rg8-z0.1} & 8 & 1 & 8 & 8& 0.1 &0.05& 2.20 & 0.61 & 4.8   \\
4 & \textsc{n16-rv1-rg8-z0.1} & 16 & 1 & 8 & 8& 0.1 &0.05& 4.60 & 1.25 & 5.2  \\
5 & \textsc{n32-rv1-rg20-z0.01}& 32 & 1 & 20 &8& 0.01 &0.05& 10.00 & 2.01 & 3.9  \\
\hline
6 & \textsc{n8-rv2-rg8-z0.1}  & 8 & 2 & 8 &8& 0.1 & 0.05&2.30 & 2.82 & 7  \\
7 & \textsc{n16-rv2-rg8-z0.1}  & 16 & 2 & 8 &8& 0.1 &0.05& 4.80 & 3.13 & 7.5   \\
8 & \textsc{n32-rv2-rg20-z0.01} & 32 & 2 & 20 & 8&0.01 &0.05& 10.30 & 3.95 & 7.2    \\
\hline
9 & \textsc{n8-rv4-rg8-z0.1}  & 8 & 4 & 8 &8& 0.1 &0.05& 2.30 & 4.66 & 11.1  \\
10 & \textsc{n16-rv4-rg8-z0.1} & 16 & 4 & 8 &8& 0.1 &0.05& 4.90 & 6.26 & 11.2   \\
\hline
11 & 47~Tuc &30  & 4 & 7.4& 4.5&0.38 &0.022 & 9.60&0.8 &7\\
12 & M22 &8 & 0.9 & 8& 3.2 &0.1 &0.05 & 2.20&1.5 &4.7 \\
\hline
\enddata
\tablecomments
{List of the \texttt{CMC Cluster Catalog} models used in this study (all are integrated to 12~Gyr) together with models tailored to fit 47~Tuc and M22 (at 10.2 and 10.9~Gyr, respectively). From left to right we give the initial number of objects $N$, initial virial radius $r_v$, Galactocentric distance $R_{\rm gc}$ (assumed constant), heliocentric distance $d$, metallicity $Z$, final total cluster mass $M_{\mathrm{tot}}$, theoretical core radius $r_c$, and half-mass radius $r_h$.
}
\end{deluxetable*}

\subsection{47 Tuc and M22} \label{47TucM22}

In addition to the models from the \texttt{CMC Cluster Catalog}, we also explore models designed to better match the specific Milky Way clusters 47~Tuc and M22. These clusters are of particular interest because they lie in front of the rich stellar backgrounds of the SMC and the Galactic bulge, respectively. We base our microlensing rate estimates for these GCs on our models of 47~Tuc \citep{Ye_2021} and M22 \citep{Kremer_2019a} that best match these clusters' observed surface brightness and velocity dispersion profiles, as determined by the $\chi^2$ fitting methodology described by \cite{Kremer_2019a} and \cite{Rui_2021}.

To match 47~Tuc, \cite{Ye_2021} vary the initial number of stars, density profile, binary fraction, virial radius, tidal radius, and IMF. The density profile of the best-fitting model is an Elson profile \cite[][]{Elson_1987} with $\gamma=2.1$. The IMF consists of a two-part power-low mass function in mass range $0.08-150 \,M_{\odot}$ with a break mass at $0.8\,M_{\odot}$ and power-law slopes of $\alpha_1 = 0.4$ and $\alpha_2 = 2.8$ \citep[][]{Giersz_2011}. The initial number of stars is $N=3\times 10^6$, with binary fraction $f_b= 0.022$ \cite[][]{Giersz_2011}, virial radius $r_v = 4 $~pc, Galactocentric distance $R_{\rm gc}= 7.4$~kpc \cite[][]{Harris_2010,Baumgardt_2019}, and metallicity $Z= 0.0038$ \citep[][2010 edition]{Harris_1996}. Following \cite{Ye_2021}, we use the snapshot at $10.2$ Gyr as a representative of the best-fit models, which span the age of $\sim 9-12$ Gyr, consistent with 47~Tuc's observationally estimated ages of $\sim 9-14$ Gyr \citep[e.g.,][and references therein]{Dotter+2010,Hansen+2013,VandenBerg+2013,Brogaard+2017,Thompson+2010,Thompson+2020}. The model observational core radius and half-light radius are 0.4~pc and 3.8~pc at $t=10.2$~Gyr, respectively.

In the case of M22, however, \cite{Kremer_2019a} investigated the effect of the initial virial radius $r_v\in [0.5,4]$~pc on the total number of BHs retained by the cluster. The best-fitting model features initial $r_v=0.9$~pc and age 10.9~Gyr. Other important initial conditions (kept fixed in the study) are initial $N=8\times 10^5$, binary fraction $f_b=0.05$, metallicity $Z=0.001$, and Galactocentric distance $R_{\rm gc}= 8$~kpc, utilizing the standard \cite{Kroupa_2001} IMF in the range $0.08$--$150\,M_{\odot}$. The model observational core radius and half-light radius are 0.9~pc and 2.4~pc at $t=10.9$~Gyr, respectively. Important initial and final parameters from our best-fitting models to 47~Tuc and M22 are also provided in Table~\ref{table:IC}. 

\section{Microlensing rates} \label{sec:rates}

We now describe how we compute the microlensing rates in our models, referring throughout to \cite{Griest_1991} and \cite{Paczynski_1996}. 

Consider a luminous background star (source) star and a faint foreground object (lens) at distances $D_S$ and $D_L$ from an observer, respectively. As they pass each other with relative proper motion $\mu_{LS}$ perpendicular to the line of sight, the lens gravitationally focuses the light from the source star, amplifying its observed brightness by a factor

\begin{equation}
A = \frac{u^2+2}{u \sqrt{u^2+4}},
\end{equation}
where $u\equiv \beta/\theta_{E}$ and $\beta$ is the angular separation between the lens and source relative to the observer. The angular Einstein radius is
\begin{equation}
\begin{aligned}
\theta_{E} &= \left(\frac{4 G m_{L}}{c^2} \frac{D_{S}-D_{L}}{D_{L} D_{S}}\right)^{1/2} \\
&\approx 92.2 \left(\frac{m_L/M_\odot}{D_L/{\rm pc}}\right)^{1/2} \left(1+\frac{D_L}{x}\right)^{-1/2} \rm{mas},
\label{einstein_radius}
\end{aligned}
\end{equation}
where $c$ is the speed of light, $m_L$ is the lens mass, and $x\equiv D_S-D_L$ is the distance between the source and lens. 
An angular separation corresponding to $u=1$ implies the source passing the lens at a projected distance equal to the Einstein radius and gives a magnification $A \approx 1.34$.
Like \cite{Zaris_2020}, we only count microlensing events that magnify the source star by more than a threshold value considered to be detectable by modern telescopes, $A_{T} = 1.01$ \citep{Bellini_2017}. The maximum allowed misalignment is then $u_{\rm max}\equiv [2A_T(A_T^2-1)^{-1/2}-2]^{1/2}\approx3.5$, corresponding to a decrease in magnitude by more than $\Delta {\rm mag} = 2.5 \log_{10} A_{T} \simeq 0.011$.

We calculate microlensing rates in our model GCs considering two different lens--source configurations. In both cases, all lenses belong to the GC and represent either stellar remnants or faint M~dwarfs in the mass range $0.08 - 0.2\,M_{\odot}$. Source stars, however, are either located inside the GC as well (the ``self-lensing'' configuration) or in a distant background system, such as the Galactic bulge or SMC (the ``background'' configuration). In principle, source stars could be anywhere and future studies may consider alternative backgrounds for other clusters. Given our specific focus on M22 and 47~Tuc, however, we only consider background sources from the Galactic bulge and SMC. For these GCs, we estimate the background microlensing rates analytically, as described in Section~\ref{sec:analytical}.
The self-lensing rates we compute numerically based on the positions and velocities of each object in our simulations, as detailed in Section~\ref{sec:numerical}.

For simplicity, we assume that all the lenses and sources are isolated (i.e., not in binaries or higher multiples). Based on the HST survey of globular clusters \cite[][]{Sarajedini_2007}, we assume that stars having masses down to about $0.2\,M_{\odot}$ can be resolved and hence act as source stars in microlensing events (at a distance of $4.5\,$kpc---the distance of 47~Tuc---this corresponds to an apparent v-band magnitude of roughly 24).
This is an optimistic approximation since observing time on telescopes capable of observing this far down the main sequence will likely remain at a premium for the foreseeable future. Additionally, in actual observations of centrally crowded clusters (like 47~Tuc), it is possible that only a few percent of $0.2\,M_{\odot}$ stars may be detected in the cluster core and a few tens of percent in the cluster halo \citep{Anderson_2008}. This is a worst-case scenario for the most centrally crowded clusters and the central completeness rapidly improves with mass and approaches $100\%$ by about $0.4\,M_{\odot}$ for many of the nearby Milky Way GCs, including M22 \citep{Weatherford_2018}. Even so, any rate estimate that we base on cluster models without correcting for observational incompleteness will over-predict the observed rate by potentially a factor of a few.

\subsection{Analytical rate estimates for distant background sources \label{sec:analytical}}
\begin{figure}
    \begin{center}
    \includegraphics[width=1\linewidth]{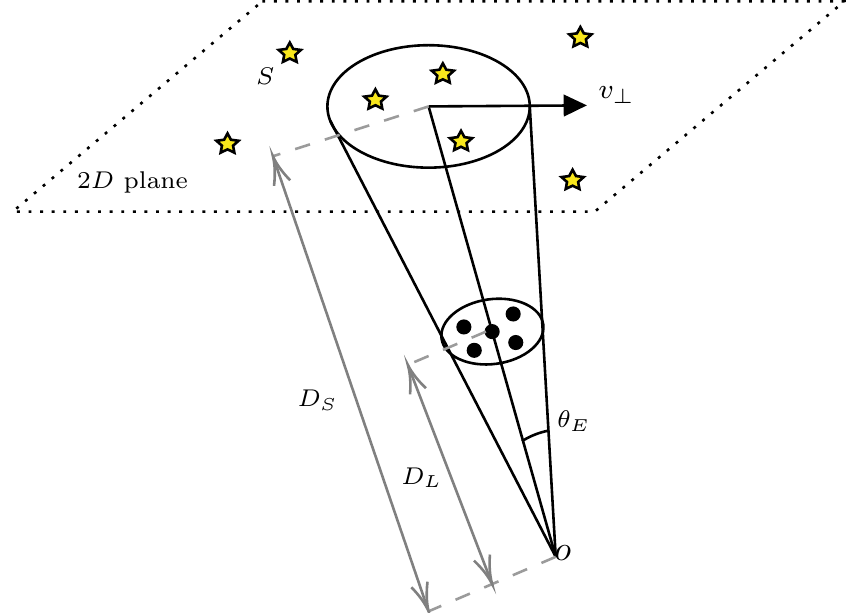}
    \caption{\footnotesize Demonstration of the lens-source geometry adopted in Section~\ref{sec:rates}. The lenses and sources are at distances $D_L$ and $D_S$, respectively, from the observer $O$ along the line of sight. All the lenses and source stars are projected onto the two-dimensional plane of the sky. To compute the microlensing rate as the lens moves along this plane, we count the number of sources that lie within the area swept out by the Einstein ring (with diameter $2R_E$) at any point during the observing time interval.}
    \label{fig:cone}
    \end{center}
\end{figure}

We now describe the method we use to compute the microlensing rates for distant background sources. To lighten the notation in all the ensuing calculations, we re-scale the angular Einstein radius $\theta_E \rightarrow u_{\rm max} \theta_E$ (characterizing microlensing events that yield magnifications $A>A_T$). The probability of observing a microlensing event meeting this criterion at any time, referred to as the optical depth, is then given by the fraction of the sky swept out by the angular area $dS$ of all lenses, that is  $d\tau = dS/\Omega$, where $\Omega$ is the solid angle of the region observed. Provided that the lens--source relative proper motion over the time interval $dt$ is approximately constant, the total rectangular area swept out by the lenses on the sky at distance $D_L$ is given by
 \begin{equation}
     dS(D_L) = (2 \theta_E \times \mu_{LS} dt) n(D_L) \Omega D_L^2 dD_L, \label{area}
 \end{equation}
where $2\theta_E \times \mu_{LS} dt$ is the angular area swept out by one lens and $n(D_L)\Omega D_L^2 dD_L$ is the number of lenses in a volume element $dV = \Omega D_L^2 dD_L$ with shell thickness $dD_L$ and lens number density $n(D_L)$ along the line of sight (see Figure~\ref{fig:cone}). Integrating the contribution of all the shells along the line of sight yields
 \cite[][]{Gaudi_2012}
 \begin{align}
    \Gamma &=  \frac{1}{\Omega}\int_0^{D_S}   2 \theta_E \mu_{LS} n(D_L) dV 
    \nonumber \\ &=  \int_0^{D_S}  2 R_E v_{\perp} n(D_L) dD_L, \label{rate}
 \end{align}
where we have used $\theta_E = R_E/D_L$ and $\mu_{LS} = v_\perp/D_L$ given the relative source-lens velocity $v_{\perp}$. Eq.~\eqref{rate} represents the probability that a single source star falls into the rectangular area of lenses \textit{per unit time}, i.e., $\Gamma = d \tau/dt$, referred to as the microlensing rate. Monitoring a total number $N_{\star}$  of stars for a duration $T_{\rm obs}$ results in the total number of microlensing events:
\begin{equation}
    N_{\rm ev} = N_{\star} \Gamma T_{\rm obs}. \label{total_event}
\end{equation}

To estimate the event rates for distant background sources, one can approximate the average lens and source distances ($D_L$ and $D_S$, respectively) as constant over the integration range in Eq.~\eqref{rate}. 
Doing so yields \cite[e.g.,][]{Fregeau_2002}
\begin{equation}
    \Gamma(r) = 2 R_E v_{\perp} \Sigma(r), \label{rate_bulge1}
\end{equation}
where $r$ is the distance perpendicular to the line of sight and $\Sigma$ is the surface density of the lenses. Eq.~\eqref{rate_bulge1} is also expressible in terms of angular quantities as follows \cite[e.g.,][]{Harding_2017}:
\begin{equation}
    \Gamma(r) = 2 \theta_E \mu_{LS} \sigma_L(r), \label{rate_bulge}
\end{equation}
where $\sigma_L$ is the surface density of the lenses in units of $\rm{arcsec}^{-2}$. Here, the angular Einstein radius and proper motion take units of arcsec and arcsec/yr, respectively.

Finally, when computing the total lens-source relative proper motions, we take the total transverse velocity to be that of typical GCs with respect to the Galactic center, i.e., $ v_{\perp} \sim 100\, \rm km\,s^{-1}$. That is, we ignore the individual motions of the lenses, typically $\sim 10\, \rm km\,s^{-1}$. 

The typical timescale over which a microlensing event takes place is obtained by 
\begin{equation}
\begin{aligned}
t_E &\equiv \frac{R_E}{v_{\perp}} = u_{\rm max} \frac{\sqrt{4G m_L D_L x/D_S}}{c v_{\perp}} \\
&\simeq 22~\mathrm{days} \left(\frac{v_{\perp}}{100~ \mathrm{km/s}}\right)^{-1} \left(\frac{m_L}{0.1 M_{\odot}}\right)^{1/2}\left( \frac{D_L x/D_S}{2~\mathrm{kpc}}\right)^{1/2},
\label{event_duration}
\end{aligned}
\end{equation}
for $u_{\rm max}$=1.

\subsection{Numerical calculations of self-lensing rates\label{sec:numerical}}
We numerically compute the self-lensing rates in our cluster models similarly to \citet{Griest_1991,Zaris_2020}. This method is more exact and accurate than the approximate analytical approach presented in the previous section but is limited to sources within the cluster (self-lensing). First we project the positions, $ r $, and tangential and radial velocities, $ v_t $ and $ v_r $, respectively, of each object onto two dimensions, with angular coordinates assigned randomly on the unit sphere. 

We imagine an Einstein ring attached to a lens moving across the corresponding sky-projected plane at a constant relative velocity, $v_{\perp}$, over an observing duration $T_{\rm obs}$ and check if the broad pass of the Einstein ring intersects with the position of a source (see Figure~\ref{fig:cone}). The area swept out on the `sky' by the lens will cover a narrow, nearly rectangular area across the plane perpendicular to the line of sight. We choose $T_{\rm obs}$ small enough  that the lens only moves a small fraction of the cluster radius (enabling the projected trajectory to reasonably be described by a straight line) but large enough that the probability it will pass through a source is non-negligible.
Finally, we compute the total microlensing rate by counting up the number of source stars inside the microlensing tube of radius $R_E =\theta_E D_S$ in the source plane and length $v_{\perp} T_{\rm obs}$ and then dividing by $T_{\rm obs}$.

\section{Results \label{sec:results}}

\subsection{The \texttt{CMC} Cluster Catalog} 

\begin{figure*}
\begin{center}
    \includegraphics[width=\textwidth]{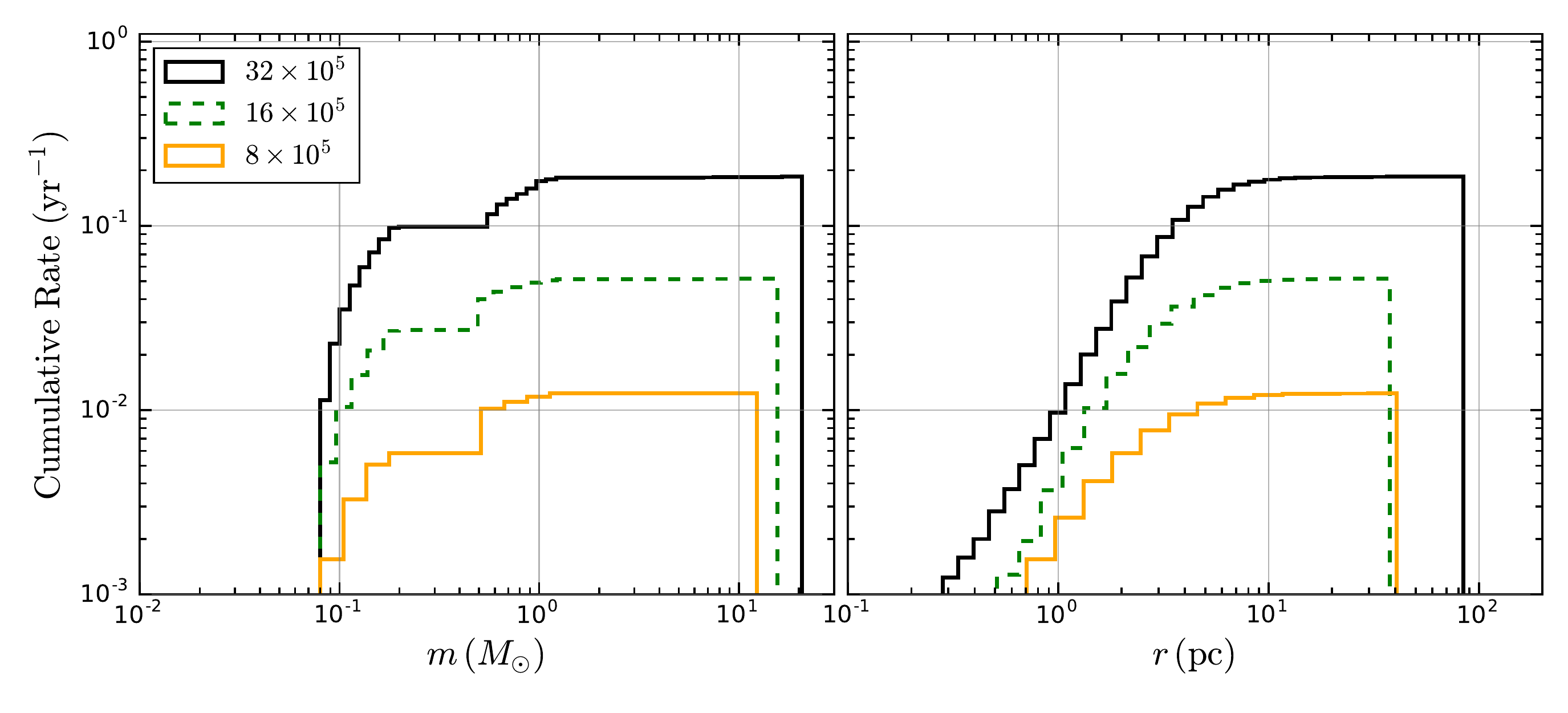}
    \caption{\footnotesize Cumulative self-lensing rates as a function of mass (left) and radial  position (right) for all lenses in three GC models with different initial total numbers of stars, namely \textsc{n8-rv1-rg8-z0.1}, \textsc{n16-rv1-rg8-z0.1} and \textsc{n32-rv1-rg20-z0.01} (see Table~\ref{table:catalog_rates}). The initial virial radius is 1~pc for all models. As the total number of stars increases, lenses span wider ranges of mass and radial distance, and the microlensing rate increases.}
\label{fig:mass_dist_catalog}
\end{center}
\end{figure*}

In this section, we present our self-lensing rate estimates for the 10 \texttt{CMC Cluster Catalog} models. Table~\ref{table:catalog_rates} lists for each model the total numbers of source stars $N_{\star}$ and various lens populations as well as the corresponding total event rates $\Gamma_{\rm tot}$ in units of events per year per $N_{\star}$. A quick glance confirms the natural expectation that the microlensing event rates are highest in clusters with the highest numbers of sources and lenses. As the numbers of both lens and source stars decrease, the microlensing event rates decrease by nearly the same factor. This trend is also evident in Figure~\ref{fig:mass_dist_catalog} showing the relative contributions from lenses of different masses and radial positions in the cluster to the total self-lensing rates for three \texttt{CMC Cluster Catalog} models with different initial $N$.
We see that the objects in the mass range $\simeq 0.1-0.7 \,M_{\odot}$, which represent WDs and M~dwarfs, dominate the microlensing rates. This is unsurprising since they dominate the cluster mass function at late times. Furthermore, as the initial $N$ increases, more NSs and BHs are retained and contribute to the microlensing rate. In Figure~\ref{fig:scatter_all}, we show the mass and radial distributions of the different types of lenses that produce an observable event in the model \textsc{n32-rv1-rg20-z0.01} with initial $N=3.2\times 10^6$ and virial radius $r_v=1\,$pc. As the lightest lenses we consider here ($\simeq 0.1\,M_{\odot}$), M~dwarfs in this massive cluster are located mostly beyond the half-mass radius (4~pc), well into the cluster halo. Among various types of WDs, carbon-oxygen WDs, with average mass $0.75\,M_{\odot}$, produce the highest microlensing rate since they dominate by number. They are massive enough to segregate toward the cluster center, but only slightly so. The same is true for the NSs, with average mass $1.3\,M_\odot$, but the BHs reside much deeper in the cluster potential given their much higher average mass ($15\,M_\odot$).

\startlongtable
\begin{deluxetable*}{l|l|ccccccccc}
\tabletypesize{\scriptsize}
\tablewidth{0pt}
\tablecaption{Self-lensing event rates of various lens populations in all model GCs\label{table:catalog_rates}}
\tablehead{
\colhead{} &
\colhead{Simulation} &
\colhead{$N_{\star}$}&
\colhead{$N_{\rm MD}$}&
\colhead{$N_{\rm WD}$}&
\colhead{$N_{\rm NS}$}&
\colhead{$N_{\rm BH}$} &
\colhead{$\Gamma_{\rm MD}$}&
\colhead{$\Gamma_{\rm WD}$}&
\colhead{$\Gamma_{\rm NS}$}&
\colhead{$\Gamma_{\rm BH}$} \\
\colhead{}&
\colhead{}&
\colhead{}&
\colhead{}&
\colhead{}&
\colhead{}&
\colhead{}&
\colhead{yr$^{-1}$}&
\colhead{yr$^{-1}$}&
\colhead{yr$^{-1}$}&
\colhead{yr$^{-1}$}
}
\startdata
1 & \textsc{n8-rv0.5-rg8-z0.1} & 274944 & 232688 & 76753 & 273 & 1  & 0.004 & 0.01 & 8$\times 10^{-5}$ & $<10^{-5}$   \\
2 & \textsc{n16-rv0.5-rg8-z0.1} & 600279 & 537756 & 161703 & 740 & 64  & 0.04 & 0.05 & 6$\times 10^{-4}$ &  2$\times 10^{-4}$   \\
\hline
3 & \textsc{n8-rv1-rg8-z0.1}& 300847 & 271497 & 80989 & 238 & 21  & 0.005 & 0.005 & 5$\times 10^{-5}$ &  2$\times 10^{-5}$   \\
4 & \textsc{n16-rv1-rg8-z0.1} & 626541 & 578016 & 164008 & 610 & 207  & 0.03 & 0.03 & 4$\times 10^{-4}$ &  3$\times 10^{-4}$   \\
5 & \textsc{n32-rv1-rg20-z0.01} & 1263364 & 1188229 & 343463 & 4901 & 962  & 0.1 & 0.1 & 0.002 & 0.003 \\
\hline
6 & \textsc{n8-rv2-rg8-z0.1}  & 314259 & 289919 & 81065 & 160 & 110  & 0.003 & 0.002 & 2$\times 10^{-5}$ &  3$\times 10^{-5}$  \\
7 & \textsc{n16-rv2-rg8-z0.1}& 640980 & 598069 & 164674 & 449 & 534  & 0.02 & 0.01 & 2$\times 10^{-4}$ &  3$\times 10^{-4}$ \\
8 & \textsc{n32-rv2-rg20-z0.01} &1286873 & 1220092 & 345888 & 4160 & 1866  & 0.08 & 0.06 & 0.001 & 0.003 \\
\hline
9 & \textsc{n8-rv4-rg8-z0.1} & 316884 & 293302 & 79556 & 74 & 297  & 0.001 & 0.001 &  $<10^{-5}$ &  2$\times 10^{-4}$  \\
10 & \textsc{n16-rv4-rg8-z0.1} & 651299 & 610092 & 164273 & 335 & 979  & 0.008 & 0.005 & 1$\times 10^{-5}$ &  3$\times 10^{-4}$ \\
\hline
11 & 47~Tuc & 1209403 & 344100 & 447959 & 1298 & 159  & 0.03 & 0.2 &  0.002 &  4$\times 10^{-4}$  \\
12 & M22 & 291315 & 259255 & 76867 & 478 & 40  & 0.004 & 0.005 & 4$\times 10^{-5}$ &  2$\times 10^{-5}$ \\
\hline
\enddata
\tablecomments
{The total number of source stars $N_{\star}$, M~dwarfs $N_{\mathrm{MD}}$, white dwarfs $N_{\mathrm{WD}}$, neutron stars $N_{\mathrm{NS}}$, and black holes $N_{\mathrm{BH}}$ at 
12, 10.2, 10.9~Gyr for all \texttt{CMC Catalog models}, 47~Tuc and M22, respectively. The microlensing rate of each lens population is obtained numerically for a given total number of source stars $N_{\star}$ in all model GCs.}
\end{deluxetable*}

\begin{figure*}
\begin{center}
    \includegraphics[width=\textwidth]{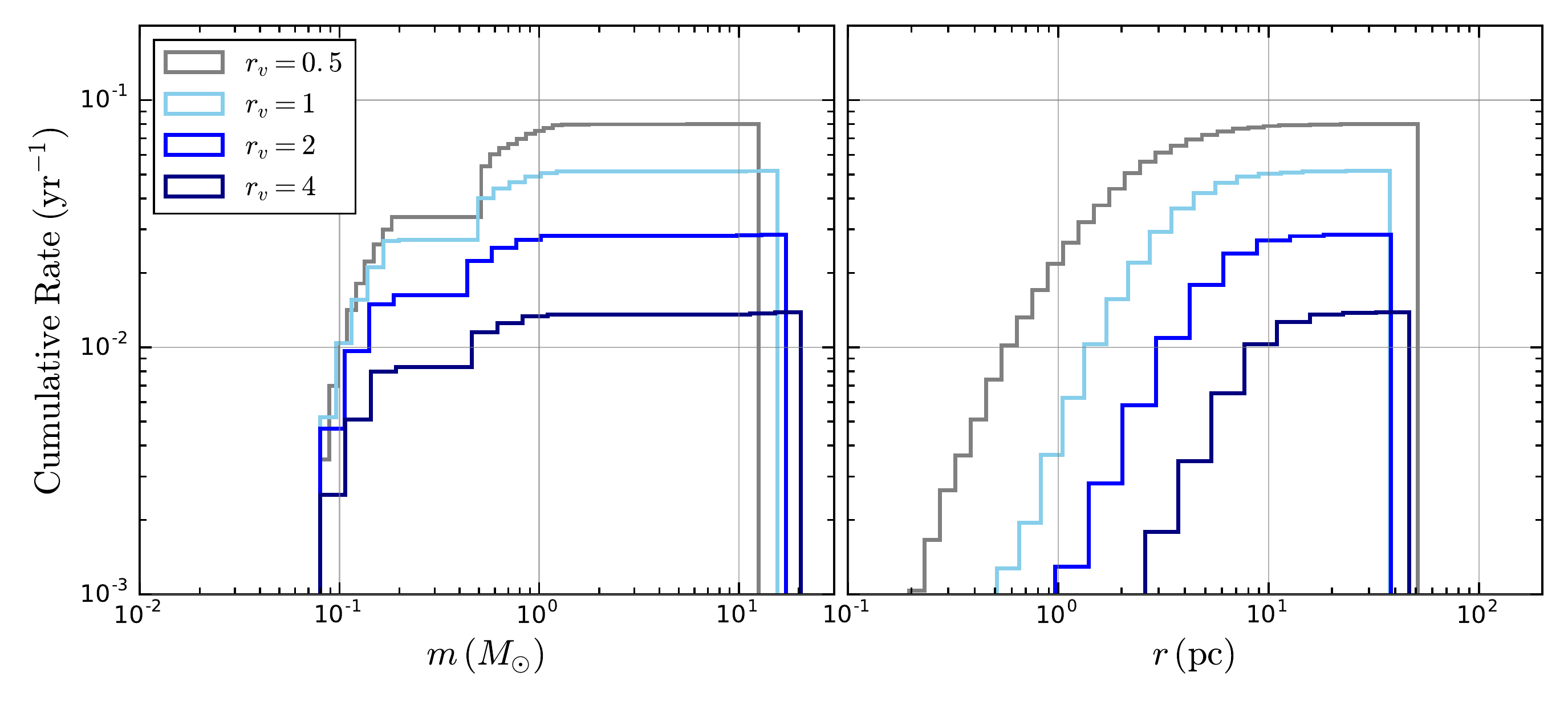}
    \caption{\footnotesize Cumulative self-lensing rates as a function of mass (left) and radial position (right) for all lenses in four GC models with different initial sizes, namely \textsc{n16-rv05-rg8-z0.1}, \textsc{n16-rv1-rg8-z0.1}, \textsc{n16-rv2-rg8-z0.1} and \textsc{n16-rv4-rg8-z0.1} (see Table~\ref{table:catalog_rates}). The initial total number of stars is $N=1.6\times 10^6$ for all models.}
\label{fig:mass_dist_catalog_virial}
\end{center}
\end{figure*}

\begin{figure*}
\begin{center}
    \includegraphics[width=0.48\textwidth]{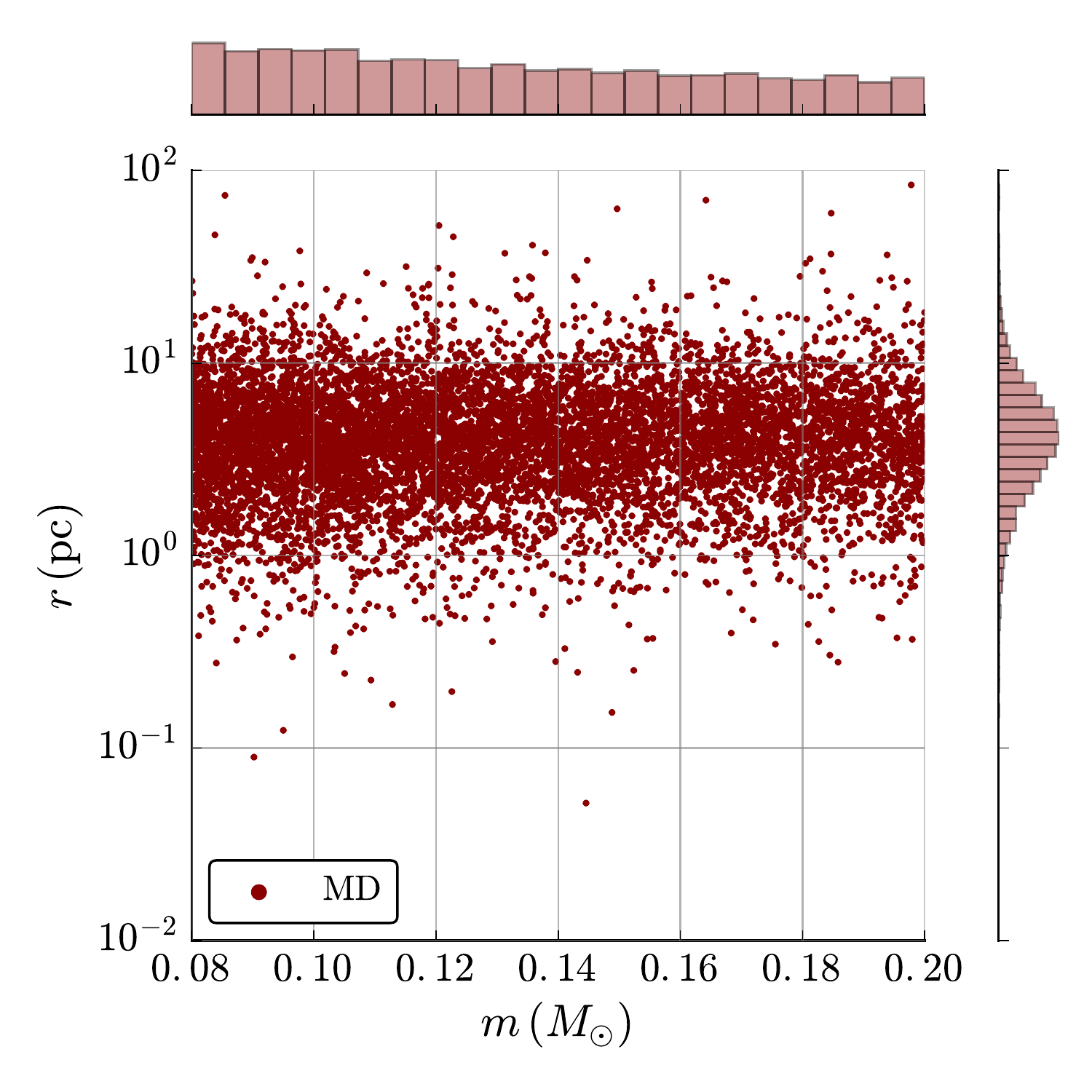}
     \includegraphics[width=0.48\textwidth]{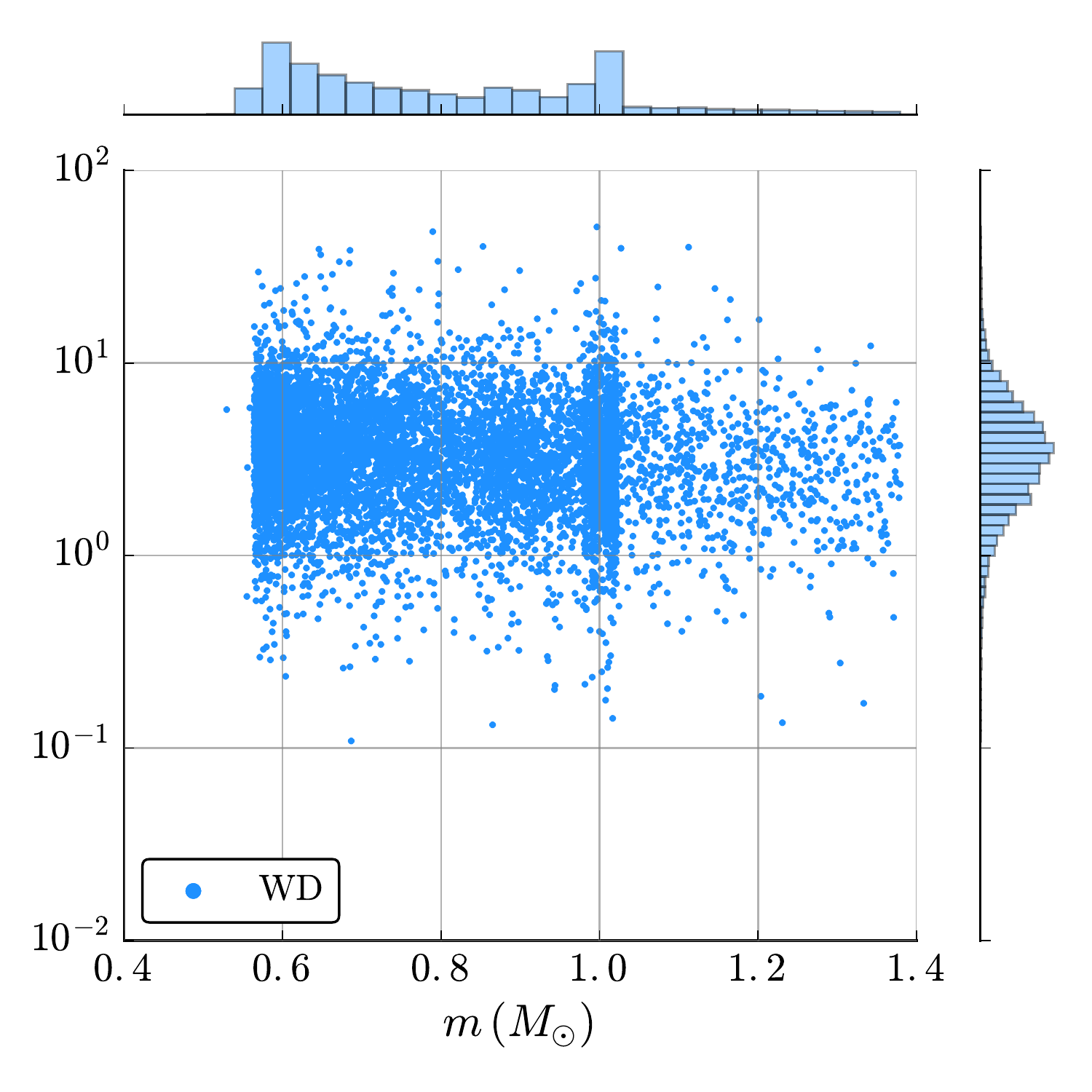}
      \includegraphics[width=0.48\textwidth]{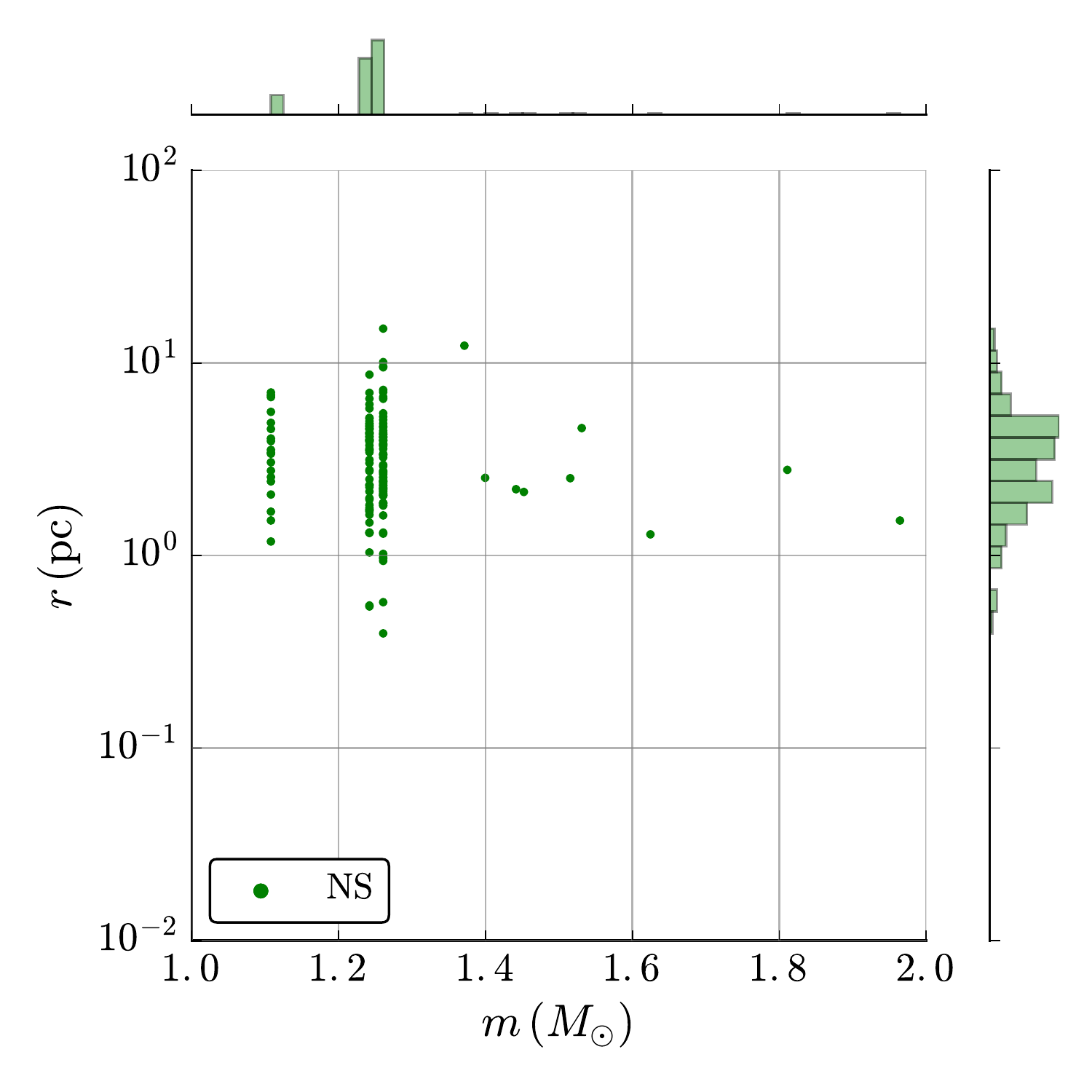}
       \includegraphics[width=0.48\textwidth]{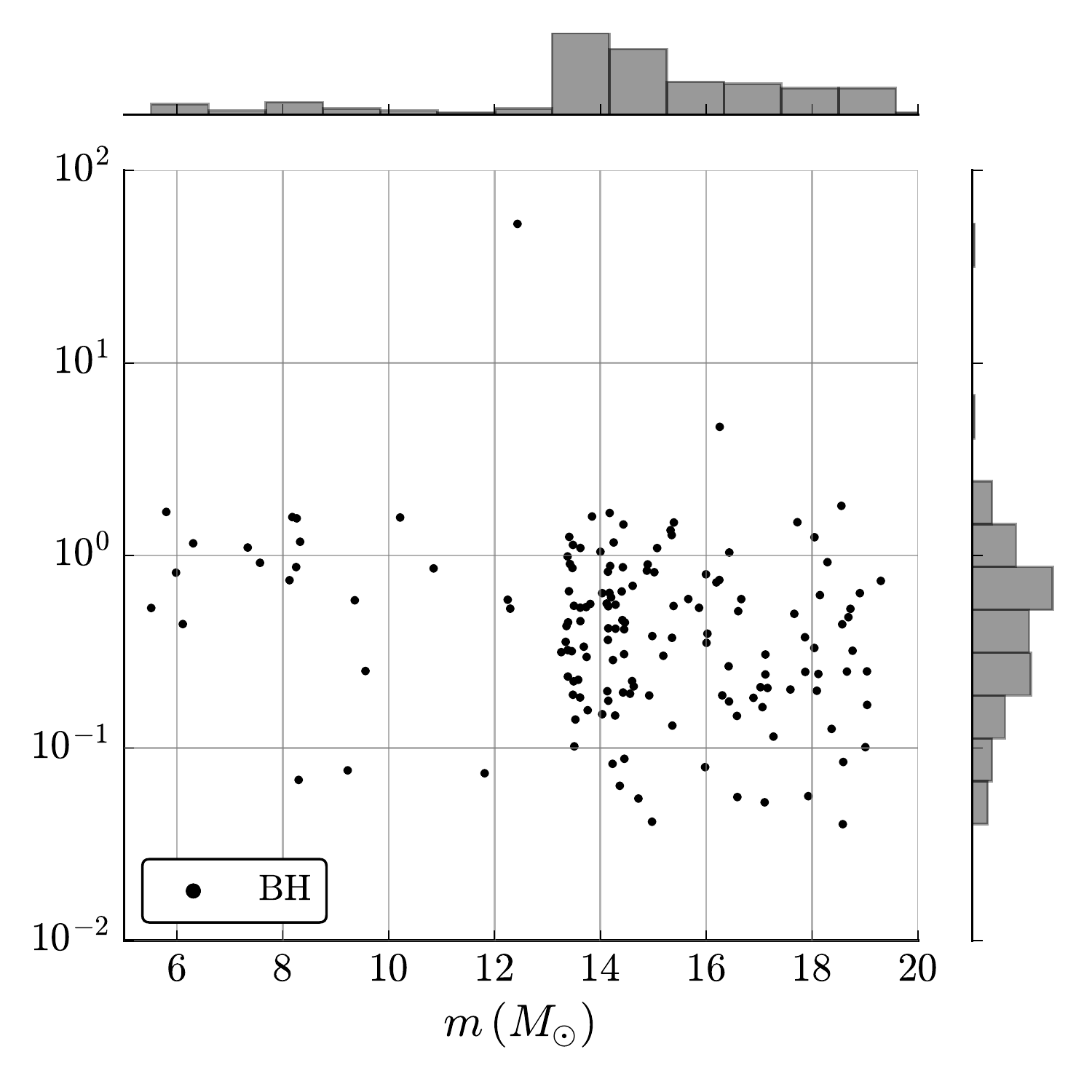}
\caption{Mass versus radial position in cluster for various lens types that caused a microlensing event in our model \textsc{n32-rv1-rg20-z0.01}, over a time interval $T_{\rm obs}\simeq 3.7\times10^4$~yr.}
\label{fig:scatter_all}
\end{center}
\end{figure*}

The initial size of the cluster also impacts microlensing rates. This impact is demonstrable in Figure~\ref{fig:mass_dist_catalog_virial} showing the cumulative distribution of the microlensing rates as functions of lens mass (left) and radial position (right) for models with different initial virial radii. These plots show that total rates increase with decreasing $r_v$ as the lenses concentrate in the inner regions of their cluster. There are more subtle contributions to this overall impact, however; $r_v$ actually influences the microlensing rates through three primary competing effects. First, the microlensing rate in Eq.~\eqref{rate} is directly proportional to the number density of lenses times the cluster radius (via the integral over distance from $D_L$ to $D_S$). Therefore the microlensing rates should naïvely scale as roughly the inverse square of the virial radius; clusters born with smaller $r_v$ should indeed exhibit higher microlensing rates. Table~\ref{table:catalog_rates} confirms this expectation for M~dwarfs, WDs, and NSs, but the tabulated results show the inverse dependence of these rates on $r_v$ is closer to linear than quadratic. This arises in part because the angular Einstein radius in Eq.~\ref{einstein_radius}---and therefore the microlensing rate in Eq.~\ref{rate}---is nearly proportional to the square-root of the source-lens distance. Since this is \emph{also} proportional to $r_v$, the average source-lens distance is smaller in denser clusters, partially negating the rate enhancement due to the higher density on its own. Finally, clusters with smaller $r_v$ dynamically evolve faster and therefore retain fewer objects (both lenses and sources) at late times than clusters born with lower density. This is especially true for BHs, as shown by \cite{Kremer_2019a}. Consequently, for a given number of source stars, the total number of BH microlensing events actually \emph{increases} slightly or remains constant with increasing virial radius.

 While BHs and BH microlensing events are depleted in dense, dynamically evolved GCs, the same logic is not applicable for WDs and NSs. In clusters old enough and centrally-dense enough to be classified as core-collapsed \citep[e.g.,][]{Trager1995}, the lack of a significant central BH population allows WDs and NSs to mass-segregate into the cluster center instead. The increase in density experienced by these compact populations can enhance WD and NS microlensing rates. For instance, among the models in Table~\ref{table:catalog_rates}, the model \textsc{n8-rv0.5-rg8-z0.1} represents a core-collapsed cluster (see Figure~5 in \cite{Kremer_2020}), which retains only one BH at $t=12$~Gyr. Given the above expectations for core-collapsed clusters, it is therefore unsurprising that this example features the lowest BH microlensing rate among the models analyzed, at less than 10 events per Myr per GC. Due to the absence of the central BHs, however, the rates for WDs and NSs are enhanced relative to the other models analyzed with the same initial $N$. In general, core-collapsed clusters should be poor candidates for BH microlensing searches, but potentially strong candidates for WD (or NS) microlensing searches.

Leaving aside the finer details of how the self-lensing rates depend on cluster initial conditions, it is clear that the rates from any single cluster remain quite small. Across the models, the self-lensing rates for BHs and NSs are of the same order and typically in the range $\Gamma \sim 10^{-5}$--$10^{-3}$ per year. However, apart from typical CMC models, massive GCs like Omega Centauri (NGC 5139) can yield substantially higher BH microlensing rates of $0.1$--$1 \, \rm{yr}^{-1}$ \citep{Zaris_2020}, assuming $10^4$ and $3\times 10^6$, BHs and visible stars in the cluster, respectively. For both WDs and M~dwarfs we find a maximum event rate $\Gamma  \sim 10^{-1} $ per year. Even for the ``best-case'' cluster model \textsc{n32-rv1-rg20-z0.01}---featuring high $N$---we find that an HST-class telescope would need to monitor 30 such GCs for about a decade to have a reasonable chance of catching a single BH self-lensing event. This rate is slightly smaller for NSs (about 50 such GCs would need to be observed). Even for WDs or M~dwarfs, it would take a year of monitoring 10 such GCs to catch a single event of each type. The key reason why the self-lensing rates are so small is that when the source and lens are both within the cluster the angular Einstein radius, and therefore the lensing cross section, becomes extremely small, i.e., $\theta_E \sim 10^{-2}$~mas for $D_S-D_L\sim 1$~pc and typical cluster distance of 8~kpc. This makes the chance of a lens--source alignment within a cluster highly unlikely. Ultimately, self-lensing events are only reasonably observable for WDs and M~dwarfs in populous clusters ($N \gtrsim 10^{6}$ stars). Otherwise, the self-lensing rates are negligible, especially for BHs and NSs. With this in mind, we continue our analysis by considering different lens--source configurations. In particular, we consider the clusters that lie in front of rich background stars, such as the Galactic bulge and the SMC.

\subsection{Best-fit models for 47~Tuc and M22}

We now present the microlensing event rates in our best-fitting models to 47~Tuc and M22, described in Section \ref{47TucM22}. In the self-lensing scenario, we estimate the rates numerically using the description explained in Section \ref{sec:numerical}. We also estimate the microlensing rates of stars in the distant backgrounds of these two GCs, the SMC (for 47~Tuc) and Galactic bulge (for M22) analytically using Eq.~\eqref{rate_bulge}, based on the average Einstein radius and surface density of each of the four lens populations in our best-fit models. The total microlensing rates i.e., $\Gamma_{\rm tot} = \Gamma N_{\star}$, are again given in units of events per year per number of observable stars $N_{\star}$. 

\subsubsection{Microlensing of distant background sources}

In Table~\ref{table:best-fit}, we give our results for the microlensing event rates of the SMC and Galactic bulge background stars by the potential lenses in 47~Tuc and M22, respectively.  Only the outermost regions of the SMC overlap with 47~Tuc so observers would realistically only be able to monitor a limited number of SMC stars for microlensing events. \cite{Kaliari_2012}, in their deep HST observations of 47~Tuc and the SMC, cover a 60~arcmin$^2$ area, revealing over $ N_{\star} = 10,000 $ SMC background stars of mass down to 0.2$\,M_{\odot}$ behind 47~Tuc. For comparison, the area covered out to the half-mass radius of 47~Tuc ($r_h=5.3$~arcmin) is about $90$~arcmin$^2$. We assume 15,000 SMC background stars uniformly distributed in area are within the half-mass radius of 47~Tuc and about 200 within the theoretical core radius of 47~Tuc ($r_c=0.6$~arcmin). In order to estimate the microlensing rates per year per $N_{\star}$ stars, we use Eq.~\eqref{rate_bulge}, setting 47~Tuc's distance to $D_L=4.5\,$kpc \cite[][]{Harris_2010} and the SMC's distance to $D_S=61\,$kpc \cite[][]{Hilditch_2005}. In addition, we take the proper motion of 47~Tuc relative to the SMC to be  $\mu_{LS}= 4.9\, \rm mas\,yr^{-1}$ \cite[][]{Anderson_2003,Tucholke_1992}. Assuming that the lenses of average mass $m_L$ located in 47~Tuc are distributed uniformly within both the core radius and the half-mass radius, we compute the surface density of lenses $\sigma_L$ in units of arcsec$^{-2}$ within these two radii and obtain the corresponding microlensing rates. 

The probability of observing a microlensing event in a GC is expected to be larger towards the center, where the density is highest. The compact object microlensing event rate should also be dominated by the most populous compact species in the cluster, i.e., WDs. 
Our results support this expectation; we find that observing a typical number of $\sim10^4$ SMC stars behind the half-mass radius of 47~Tuc yield a microlensing rate $\Gamma \sim 1$ per year owing to WDs but $\Gamma \sim 10^{-3}$ per year owing to NSs and BHs.
The total microlensing rates depend on number of background sources, hence the survey area. Therefore, increasing the survey radius from the core to the half-mass radius increases the chances of observing microlensing events.

At distance $D_L=3.2$~kpc \cite[][]{Monaco_2004,Harris_2010}, M22 is closer to Earth and resides in front of the Galactic bulge, where one can monitor $\sim10^5$ background stars as targets \cite[e.g.,][]{Sahu_2001}. In our calculations, we set the Galactic bulge's distance to be $D_S=8.5$~kpc \cite[][]{Eisenhauer_2003,Gillessen_2009,Vanhollebeke_2009} and M22's proper motion relative to the bulge to be $\mu_{LS}= 12.2~ \rm mas\,yr^{-1}$ \cite[e.g.,][]{Bellini_2014,Kains_2016}. Similarly to the case of 47~Tuc, we find that observing a typical number of $N_{\star}=10^5$ bulge source stars \cite[e.g.,][]{Sahu_2001} within the half-mass radius ($r_h=5$~arcmin) of M22 for a decade would yield a microlensing rate $\Gamma \gtrsim 10$ per pear owing to WDs and $\Gamma \gtrsim 10^{-2}$ per year owing to NSs and BHs. The total number of microlensing events is about 10 times larger in the M22-bulge case than the 47~Tuc-SMC case thanks to the higher number of observable stars in the Galactic bulge and higher proper motion of M22.

We also analytically estimate the potential microlensing rates due to hypothetical intermediate-mass black holes (IMBHs) with masses $M_{\rm BH} \sim 10^2$--$10^4\,M_\odot$ at the centers of 47~Tuc and M22, assuming several different IMBH masses. The literature on the potential presence of IMBHs in GCs is extensive; possible detections have been suggested in several Milky Way GCs, including 47~Tuc \citep{Kiziltan_2017}, but are still debated at the light of constraints imposed by radio surveys \citep{Strader_2012b,Tremou_2018} and/or numerical modeling \citep[e.g.,][]{Henault-Brunet_2020}. Overall, the evidence seems to suggest that IMBHs larger than a few hundreds solar masses may be rare in Milky Way GCs.
In the case of the IMBH microlensing rate of background stars, we calculate the optical depth in a slightly different way. Unlike the total area in Eq.~\eqref{area}, we consider the total area covered by background stars on the sky and calculate the probability of a background star to reside inside the Einstein ring of the IMBH. Integrating over the source distances gives
\begin{equation}
    \Gamma_{\rm IMBH}(r) = 2 \theta_E \mu_{LS} \sigma_S(r), \label{rate_IMBH}
\end{equation}
where $\sigma_S$ is the surface density of the sources in units of $\rm{arcsec}^{-2}$. To estimate  the microlensing rates, we then employ Eq.~\eqref{rate_IMBH}, taking the number density of background stars to be $0.02 \, \rm arcsec^{-2}$ in the SMC  and $1.3 \, \rm arcsec^{-2} $ in the Galactic bulge \citep{Bellini_2014}. The mass of the hypothetical IMBH is varied between $10^2$--$10^4 \,M_{\odot}$ and the corresponding microlensing rates scale with the square root of the IMBH mass, $\Gamma~\sim \sqrt{m_L}$. While we find a maximum IMBH event rate of the order $10^{-4}$~yr$^{-1}$ for the 47~Tuc-SMC configuration, the M22-Galactic bulge event rates are 100 times larger $\Gamma~\sim 10^{-2}$~yr$^{-1}$ thanks to the high density of the bulge stars behind M22. Largely for this reason, \cite{Kains_2016} recently determined that M22 is the Galactic GC with the highest chance of yielding an IMBH detection via astrometric microlensing, if an IMBH were to be present in the cluster. Followup analysis of archival data from HST, however, found no evidence for astrometric microlensing in M22 \citep{Kains_2018}. Despite this preliminary analysis, M22 remains one of the best GC candidates for a microlensing survey.

\startlongtable
\begin{deluxetable*}{l|ccccc|cc}
\tabletypesize{\scriptsize}
\tablewidth{0pt}
\tablecaption{Microlensing event rates for 47~Tuc and M22
\label{table:best-fit}}
\tablehead{
	\colhead{Lens} &
	\colhead{Total number} &
    \colhead{$m_L$} &
    \colhead{$\theta_E$} &
	\multicolumn{2}{c}{Lens Density} &
	\multicolumn{2}{c}{Total Rate}\\
	\colhead{} &
	\colhead{} &
	\colhead{$(M_{\odot}$)} &
	\colhead{$(\rm{mas})$} &
	\multicolumn{2}{c}{per arcsec$^2$} &
	\multicolumn{2}{c}{per year}
	} 
\startdata
47~Tuc-SMC (t=10.2~Gyr) & &  &  &$r<r_c$ &$r<r_h$  &$r<r_c$ &$r<r_h$\\
\hline
M~dwarfs &344100 & 0.14 & 1.7  & 1.8 & 1.4 & 0.006 &0.3  \\
All White Dwarfs &447959 & 0.66 & 3.8  & 23 & 2.9 & 0.18 & 1.5 \\
He White Dwarfs & 368& 0.21 & 2.1 & 0.01  &0.0023 & 0.00004&0.00075   \\
CO White Dwarfs  &437958 & 0.65 & 3.7 &  21.3 & 2.8& 0.16& 1.5   \\
ONe White Dwarfs  & 963& 1.2 & 5.1 &1.7   &0.07 &0.017 &0.054    \\
Neutron Stars &1298 & 1.26 & 5.2  & 0.3 & 0.01 & 0.0034 &0.009 \\
Black Holes &159 & 13 &  16.7  & 0.08 & 0.001 & 0.0026 & 0.003\\
\hline
M22-Bulge (t=10.9~Gyr) &  & & &$r<r_c$ &$r<r_h$  &$r<r_c$ &$r<r_h$\\
\hline
M~dwarfs &259255 & 0.13 & 1.6 & 0.9 & 1 &0.4 &4   \\
All White Dwarfs &76867 & 0.72 & 3.8  & 1.5 & 0.6 &1.4 &6  \\
He White Dwarfs &35 & 0.24 & 2.2 &  0.0006 & 0.00014&0.0003 &0.0008  \\
CO White Dwarfs &74881 & 0.7 & 3.8 & 1.4  &0.6 &1.3 & 6 \\
ONe White Dwarfs &1951 & 1.3 & 5.1 & 0.09 & 0.02 & 0.1& 0.26  \\
Neutron Stars &478 & 1.25 &5& 0.02 & 0.005& 0.03 &0.06 \\
Black Holes &40 & 12 & 15.8  &0.005  &  0.0005& 0.02 &0.02  \\
\hline
\enddata
\tablecomments{Total microlensing event rates in 47~Tuc (monitoring $10^4$ background stars near SMC) and M22 (monitoring $10^5$ background stars in the MW bulge), as well as the total number of lenses, average lens mass  $m_L$, average Einstein radius $\theta_E $ (already scaled by $u_{\rm max}=3.5$), surface density of lenses $\Sigma$ within the core radius, and the half-mass radius of each cluster at final time 10.2~Gyr and 10.9~Gyr, respectively. We adopt the values 4.9 and 12.2~$\rm{mas}~\rm{yr}^{-1}$ for the total proper motions of 47~Tuc and M22, respectively. The half-mass $r_h$ and theoretical core radius $r_c$ are taken to be 7~pc and 0.8~pc for 47~Tuc, and 4.7~pc and 1.5~pc for M22, respectively, which are listed on Table \ref{table:IC}. We assume a number of 15,000 and 200 background stars can be observed within the half-mass and core radius of 47~Tuc, respectively. For M22, we assume a number of 100,000 and 10,000 background stars can be observed within its half-mass and core radius, respectively. These microlensing rates are obtained analytically using Eq.~\eqref{rate_bulge}.}
\end{deluxetable*}

\subsubsection{Self-lensing rates}
Finally, we summarize our numerical estimates for the self-lensing rates in 47~Tuc and M22, already quoted in Table~\ref{table:catalog_rates}. This time the Einstein radius of a lens located in either cluster is about 100 times smaller than when monitoring background stars in the Galactic bulge or SMC. The individual stellar motions ($\simeq 0.4\,\rm mas\,yr^{-1}$) relative to the clusters' centers-of-mass are also about 10 times smaller than that of the clusters' bulk motions through the Galaxy. In the self-lensing case of M22, these two factors cause the total microlensing rate to be about 1000 times smaller than the microlensing rate due to the cluster's background in the Galactic bulge, that is $\Gamma_{\rm self}~\sim 10^{-2}$ per year. For 47~Tuc, however, one can in principle monitor all $N_{\star}=10^6$ potential source stars within the cluster itself, 100 times more than the number of observable SMC stars in the cluster's background. So for 47~Tuc, the self-lensing rates are instead $\sim 10$ times smaller than the microlensing rate of the cluster's SMC background. Although this is a lower reduction than for M22, the total self-lensing rate remains rather unpromising for either cluster.

\begin{figure*}
    \begin{center}
    \includegraphics[width=\textwidth]{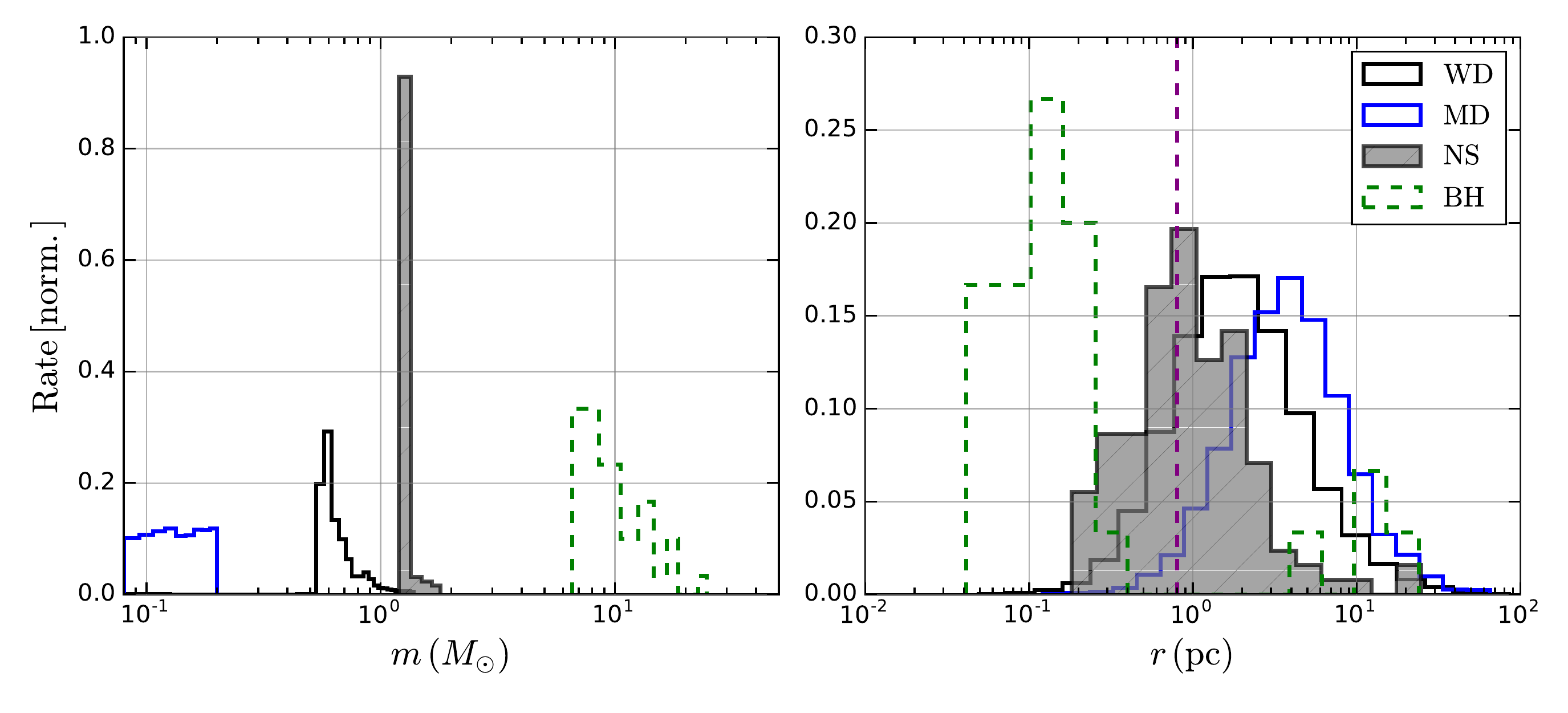}
    \caption{\footnotesize Mass (left) and radial (right) distributions of various types of lenses that caused a self-lensing event in 47~Tuc. The purple vertical dashed line (right panel) indicates the theoretical core radius.}
    \label{fig:47Tuc_plots}
    \end{center}
\end{figure*}

For completeness we show in Figure~\ref{fig:47Tuc_plots} the mass and radial distributions of different lens populations causing self-lensing events in our best-fitting model for 47~Tuc. It can be seen that while equally-spaced logarithmic mass bins of M~dwarfs between $0.08$--$0.2 \,M_{\odot}$ contribute to the event rates nearly equally, the event rates from different mass bins of WDs differ a lot. Carbon-oxygen WDs, the most numerous WD species with masses roughly between $0.5$--$1.2\,M_{\odot}$, naturally dominate the WD self-lensing rate. A small number of massive oxygen-neon WDs also contribute to the rates. However, there are very few microlensing events of helium WDs because of their relative rarity. Naturally the NS mass range is narrow but we see in the right-hand panel that self-lensing events from NSs on average occur near the theoretical core radius of the cluster (0.8~pc). The mass distribution of BHs causing self-lensing events ranges from $7$--$25\,M_{\odot}$, with a median $\simeq 10 \,M_{\odot}$, about of the overall median BH mass in our models. On average, BH self-lensing events occur at a distance of about 0.1~pc from the center of our best-fitting 47~Tuc model.

\section{Discussion and Conclusions} \label{sec:discussion}

Motivated by the idea that the high stellar densities in GCs present ideal opportunities for microlensing detections, we examine the microlensing rates observers may expect from evolved GCs based on models from the \texttt{CMC Cluster Catalog}. NSs and BHs have relatively larger lensing cross sections than WDs because the Einstein radius scales with the square root of the lens mass. Because they are much more common in dynamically evolved clusters, however, we find WDs yield the highest microlensing rates among the compact object species. The microlensing rates for M~Dwarfs ($m\simeq 0.1\,M_{\odot}$) are similar to the rates for WDs because although the WD lensing cross sections are about three times larger---i.e., $ (m_{\rm WD}/m_{\rm MD})^{1/2} \simeq 3 $---M~Dwarfs outnumber WDs in GCs by the same amount. In general, we find that the microlensing event rate is linearly proportional to the numbers of lenses and source stars and scales inversely with the cluster's virial radius. (BHs are an important exception because they are ejected more rapidly in denser clusters). For this reason, our results can be easily scaled to estimate microlensing rates for other, similar clusters, which may have different numbers of lenses or source stars.

When considering only ``self-lensing'' events between lenses and source stars that are both within the cluster we find that the microlensing rate is negligible. The BH and NS self-lensing rates in our cluster models are of order $10^{-3}$ events per year or less.  While WD and M~dwarf self-lensing rates ($\Gamma \sim 10^{-2}$--$10^{-1}$ per year) are somewhat more promising for highly populated clusters ($N \sim 10^6$) with small initial virial radii ($r_v=0.5$~pc), the rates are still very low considering practical observing timescales. However, that the rates are highest in such centrally dense, core-collapsed clusters is still an important guidepost for any potential microlensing surveys of GCs. Our recent numerical modeling of NGC~6397 \citep[][]{Kremer_2021} shows that massive WDs likely govern the dynamics of core-collapsed clusters' innermost regions ($\lesssim 0.07$ pc) and naturally account for NGC~6397's dark central population reported by \cite{Vitral_2021}. This suggests that microlensing surveys of core-collapsed Milky Way GCs may help reveal large populations of WDs, especially since their denser cores relative to non-core-collapsed clusters provide higher microlensing probabilities.

Crucially, however, self-lensing events are only part of the story. Our estimates of the total microlensing event rates in 47~Tuc and M22 demonstrate that microlensing events are far more likely to be detected from source stars in these clusters' backgrounds---the SMC and Galactic bulge, respectively---than from sources within the clusters themselves. This is mainly due to the larger microlensing cross sections (higher source-lens distance) and higher bulk proper motions of a GC relative to distant stellar fields, compared to its internal velocity dispersion. We find that total observable microlensing events of BHs and NSs should occur at similar rates of $10^{-3}$--$10^{-1}$ per year per GC when monitoring the backgrounds of these clusters, at least two orders of magnitude higher than the rate provided by self-lensing alone. WD and M~dwarf background microlensing events are a further 100 times more frequent, i.e., $\Gamma \sim 1$--$10$ per year per GC for clusters set against richly populated backgrounds like Galactic bulge or SMC. The number of WDs in GCs estimated from observations roughly matches the number of WDs in our models \citep{Cool_1996,Richer_1995,Richer_1997}, indicating these rate predictions are reasonably realistic. Such WDs would have an average mass $\simeq 0.7\,M_{\odot}$ and would mostly lie within the cluster's central few parsecs. Indeed, microlensing searches may be uniquely helpful for accessing isolated, dim WDs in the crowded centers of GCs, which otherwise would be hard to identify among brighter stars.

Here we also report the timescale of lensing events using Eq.~\eqref{event_duration} with the motivation that the differences in these characteristic timescales could help to distinguish between events caused by different species. In the self-lensing case, assuming a typical observing distance of 10~kpc, a typical GC velocity dispersion of $10\,{\rm km}\,{\rm s}^{-1}$, and a magnification threshold $A_T=1.01$ with $u_{\rm max}=3.5$, the microlensing events are expected to last about 17, 54 and 152 days for lens masses 0.1, 1 and 10$\,M_{\odot}$, respectively. On the other hand, in the case of a distant background star at $D_S \simeq 50~$kpc, the lensing duration is $\sim$~10 times longer than that of self-lensing events due to higher lensing cross sections at higher lens-source distances. 
In case of the detection of any microlensing event, one initial observation per $t_E$ is sufficient, though follow-up observations could be required to characterize the source. The lensing timescales can also be used to obtain lens masses via Eq.~\eqref{event_duration} as long as the transverse velocity and lens-source distances are known. Otherwise, one can combine photometric and astrometric analysis of microlensing events to constrain the lens mass \citep[e.g.,][]{Sahu_2022,Rybicki_2022,Lam_2022}.

In summary, microlensing searches can help address limitations of more common observational methods and extend compact object observations to isolated objects without a binary companion. In the context of these searches, many studies have been performed to constrain the mass function of compact objects using observed photometric and astrometric microlensing events \citep{Gould_2000,Wyrzykowski_2016,Sahu_2017,Ofek_2018,McGill_2019,Wyrzykowski_2020,Mroz_2021}. With the increasing number of microlensing surveys and improved monitoring of high stellar density regions like GCs, it should also be possible to obtain detailed information on the masses and kinematics of compact object populations in GCs. In particular, the upcoming survey of the Vera C. Rubin observatory \citep[][]{LSST_2009} may be ideal for lensing detections in GCs as it will look at the Milky Way GCs for ten years. By monitoring a number of $\sim 10^{10}$ stars over 10 yr, the Legacy Survey of Space and Time (LSST) is expected to detect thousands of microlensing events in the galactic field \citep{Drlica_2019}. According to our results, the Rubin Observatory may detect up to 3 (40) M dwarf lensing events and  15 (60) WD lensing events for 47 Tuc (M22) ``for free'' as part of their large-area surveys, over a 10-yr observing campaign. Additionally, we predict Rubin may observe at most $\mathcal{O}(1)$ BH and NS lensing events in the Milky Way GCs over its lifetime.

These searches might also enable detection of very low-mass objects such as brown dwarfs and free-floating planets residing in the halos of GCs. Although very poorly constrained observationally, these low-mass objects may in principle be quite numerous in GCs \citep[][]{Fregeau_2002}. The null results from photometric surveys searching for transiting giant planets  in open clusters \citep[][]{VanSaders_2011} and globular clusters \citep[][]{Gilliland_2000,Weldrake_2008} may indicate that most of them are free-floating, rather than bound to stars, and thus may in principle serve as viable lenses \citep[][]{Sumi_2011,DiStefano_2012}. Including large populations of low-mass objects within \texttt{CMC} models is beyond the scope of this paper. In future work, however, we hope to generalize \texttt{CMC} models to include more detailed treatments of planets and brown dwarfs.

\begin{acknowledgments}
This work was supported by NSF Grants AST-1716762 and AST-2108624 at Northwestern University and through the computational resources and staff contributions provided for the Quest high performance computing facility at Northwestern University. Quest is jointly supported by the Office of the Provost, the Office for Research, and Northwestern University Information Technology. KK is supported by an NSF Astronomy and Astrophysics Postdoctoral Fellowship under award AST-2001751. N.C.W. acknowledges support from the CIERA Riedel Family Graduate Fellowship. F.K acknowledges support from the CIERA BoV Graduate Fellowship. G.F., N.C.W., and F.A.R. acknowledge support from NASA Grant 80NSSC21K1722. \newline
\software{\texttt{CMC} \citep{Joshi_2000,Joshi_2001,Fregeau_2003, Fregeau_2007, Chatterjee_2010,Chatterjee_2013b,Pattabiraman_2013,Umbreit_2012,Morscher_2015,Rodriguez+2016million,Rodriguez2021}, \texttt{Fewbody} \citep{fregeau2004stellar}, \texttt{COSMIC} \citep{Breivik_2020}, \texttt{NumPy} \citep{2020NumPy-Array}, \texttt{matplotlib} \citep{Hunter2007}, \texttt{pandas} \citep{mckinney-proc-scipy-2010,jeff_reback_2021_4681666}.}
\end{acknowledgments}
\bibliography{refs}

\begin{thebibliography}{}
\expandafter\ifx\csname natexlab\endcsname\relax\def\natexlab#1{#1}\fi
\providecommand{\url}[1]{\href{#1}{#1}}
\providecommand{\dodoi}[1]{doi:~\href{http://doi.org/#1}{\nolinkurl{#1}}}
\providecommand{\doeprint}[1]{\href{http://ascl.net/#1}{\nolinkurl{http://ascl.net/#1}}}
\providecommand{\doarXiv}[1]{\href{https://arxiv.org/abs/#1}{\nolinkurl{https://arxiv.org/abs/#1}}}

\bibitem[{Alcock {et~al.}(1995)Alcock, Allsman, Axelrod, Bennett, Cook,
  Freeman, Griest, Guern, Lehner, Marshall, Park, Perlmutter, Peterson, Pratt,
  Quinn, Rodgers, Stubbs, \& Sutherland}]{Alcock_1995}
Alcock, C., Allsman, R.~A., Axelrod, T.~S., {et~al.} 1995, Phys. Rev. Lett.,
  74, 2867, \dodoi{10.1103/PhysRevLett.74.2867}

\bibitem[{{Alcock} {et~al.}(1996){Alcock}, {Allsman}, {Axelrod}, {Bennett},
  {Cook}, {Freeman}, {Griest}, {Guern}, {Lehner}, {Marshall}, {Park},
  {Perlmutter}, {Peterson}, {Pratt}, {Quinn}, {Rodgers}, {Stubbs}, \&
  {Sutherland}}]{Alcock_1996}
{Alcock}, C., {Allsman}, R.~A., {Axelrod}, T.~S., {et~al.} 1996, \apj, 461, 84,
  \dodoi{10.1086/177039}

\bibitem[{{Alcock} {et~al.}(2000){Alcock}, {Allsman}, {Alves}, {Axelrod},
  {Becker}, {Bennett}, {Cook}, {Dalal}, {Drake}, {Freeman}, {Geha}, {Griest},
  {Lehner}, {Marshall}, {Minniti}, {Nelson}, {Peterson}, {Popowski}, {Pratt},
  {Quinn}, {Stubbs}, {Sutherland}, {Tomaney}, {Vandehei}, \&
  {Welch}}]{Alcock_2000}
{Alcock}, C., {Allsman}, R.~A., {Alves}, D.~R., {et~al.} 2000, \apj, 542, 281,
  \dodoi{10.1086/309512}

\bibitem[{Anderson \& King(2003)}]{Anderson_2003}
Anderson, J., \& King, I.~R. 2003, \aj, 126, 772, \dodoi{10.1086/376480}

\bibitem[{{Anderson} {et~al.}(2008){Anderson}, {Sarajedini}, {Bedin}, {King},
  {Piotto}, {Reid}, {Siegel}, {Majewski}, {Paust}, {Aparicio}, {Milone},
  {Chaboyer}, \& {Rosenberg}}]{Anderson_2008}
{Anderson}, J., {Sarajedini}, A., {Bedin}, L.~R., {et~al.} 2008, \aj, 135,
  2055, \dodoi{10.1088/0004-6256/135/6/2055}

\bibitem[{Antonini \& Gieles(2020)}]{Antonini_2020}
Antonini, F., \& Gieles, M. 2020, \mnras, 492, 2936,
  \dodoi{10.1093/mnras/stz3584}

\bibitem[{Askar {et~al.}(2018)Askar, Arca~Sedda, \& Giersz}]{Askar_2018}
Askar, A., Arca~Sedda, M., \& Giersz, M. 2018, \mnras, 478, 1844,
  \dodoi{10.1093/mnras/sty1186}

\bibitem[{Askar {et~al.}(2016)Askar, Szkudlarek, Gondek-Rosińska, Giersz, \&
  Bulik}]{Askar_2016}
Askar, A., Szkudlarek, M., Gondek-Rosińska, D., Giersz, M., \& Bulik, T. 2016,
  \mnras, 464, L36, \dodoi{10.1093/mnrasl/slw177}

\bibitem[{Bae {et~al.}(2014)Bae, Kim, \& Lee}]{Bae_2014}
Bae, Y.-B., Kim, C., \& Lee, H.~M. 2014, \mnras, 440, 2714,
  \dodoi{10.1093/mnras/stu381}

\bibitem[{Banerjee(2017)}]{Banerjee_2017}
Banerjee, S. 2017, \mnras, 467, 524, \dodoi{10.1093/mnras/stw3392}

\bibitem[{Banerjee {et~al.}(2010)Banerjee, Baumgardt, \&
  Kroupa}]{Banerjee_2010}
Banerjee, S., Baumgardt, H., \& Kroupa, P. 2010, \mnras, 402, 371,
  \dodoi{10.1111/j.1365-2966.2009.15880.x}

\bibitem[{Baumgardt {et~al.}(2018)Baumgardt, Hilker, Sollima, \&
  Bellini}]{Baumgardt_2019}
Baumgardt, H., Hilker, M., Sollima, A., \& Bellini, A. 2018, \mnras, 482, 5138,
  \dodoi{10.1093/mnras/sty2997}

\bibitem[{Bellini {et~al.}(2017)Bellini, Anderson, Bedin, King, van~der Marel,
  Piotto, \& Cool}]{Bellini_2017}
Bellini, A., Anderson, J., Bedin, L.~R., {et~al.} 2017, \apj, 842, 6,
  \dodoi{10.3847/1538-4357/aa7059}

\bibitem[{{Bellini} {et~al.}(2014){Bellini}, {Anderson}, {van der Marel},
  {Watkins}, {King}, {Bianchini}, {Chanam{\'e}}, {Chandar}, {Cool}, {Ferraro},
  {Ford}, \& {Massari}}]{Bellini_2014}
{Bellini}, A., {Anderson}, J., {van der Marel}, R.~P., {et~al.} 2014, \apj,
  797, 115, \dodoi{10.1088/0004-637X/797/2/115}

\bibitem[{{Bond} {et~al.}(2001){Bond}, {Abe}, {Dodd}, {Hearnshaw}, {Honda},
  {Jugaku}, {Kilmartin}, {Marles}, {Masuda}, {Matsubara}, {Muraki}, {Nakamura},
  {Nankivell}, {Noda}, {Noguchi}, {Ohnishi}, {Rattenbury}, {Reid}, {Saito},
  {Sato}, {Sekiguchi}, {Skuljan}, {Sullivan}, {Sumi}, {Takeuti}, {Watase},
  {Wilkinson}, {Yamada}, {Yanagisawa}, \& {Yock}}]{Bond_2001}
{Bond}, I.~A., {Abe}, F., {Dodd}, R.~J., {et~al.} 2001, \mnras, 327, 868,
  \dodoi{10.1046/j.1365-8711.2001.04776.x}

\bibitem[{Breen \& Heggie(2013)}]{Breen_2013}
Breen, P.~G., \& Heggie, D.~C. 2013, \mnras, 432, 2779,
  \dodoi{10.1093/mnras/stt628}

\bibitem[{Breivik {et~al.}(2020)Breivik, Coughlin, Zevin, Rodriguez, Kremer,
  Ye, Andrews, Kurkowski, Digman, Larson, \& Rasio}]{Breivik_2020}
Breivik, K., Coughlin, S., Zevin, M., {et~al.} 2020, \apj, 898, 71,
  \dodoi{10.3847/1538-4357/ab9d85}

\bibitem[{{Brogaard} {et~al.}(2017){Brogaard}, {VandenBerg}, {Bedin}, {Milone},
  {Thygesen}, \& {Grundahl}}]{Brogaard+2017}
{Brogaard}, K., {VandenBerg}, D.~A., {Bedin}, L.~R., {et~al.} 2017, \mnras,
  468, 645, \dodoi{10.1093/mnras/stx378}

\bibitem[{{Camilo} \& {Rasio}(2005)}]{Camilo_2005}
{Camilo}, F., \& {Rasio}, F.~A. 2005, in Astronomical Society of the Pacific
  Conference Series, Vol. 328, Binary Radio Pulsars, ed. F.~A. {Rasio} \& I.~H.
  {Stairs}, 147.
\newblock \doarXiv{astro-ph/0501226}

\bibitem[{{Casertano} \& {Hut}(1985)}]{Casertano_1985}
{Casertano}, S., \& {Hut}, P. 1985, \apj, 298, 80, \dodoi{10.1086/163589}

\bibitem[{{Chatterjee} {et~al.}(2010){Chatterjee}, {Fregeau}, {Umbreit}, \&
  {Rasio}}]{Chatterjee_2010}
{Chatterjee}, S., {Fregeau}, J.~M., {Umbreit}, S., \& {Rasio}, F.~A. 2010,
  \apj, 719, 915, \dodoi{10.1088/0004-637X/719/1/915}

\bibitem[{Chatterjee {et~al.}(2017{\natexlab{a}})Chatterjee, Rodriguez,
  Kalogera, \& Rasio}]{Chatterjee_2017a}
Chatterjee, S., Rodriguez, C.~L., Kalogera, V., \& Rasio, F.~A.
  2017{\natexlab{a}}, \apjl, 836, L26, \dodoi{10.3847/2041-8213/aa5caa}

\bibitem[{Chatterjee {et~al.}(2017{\natexlab{b}})Chatterjee, Rodriguez, \&
  Rasio}]{Chatterjee_2017b}
Chatterjee, S., Rodriguez, C.~L., \& Rasio, F.~A. 2017{\natexlab{b}}, \apj,
  834, 68, \dodoi{10.3847/1538-4357/834/1/68}

\bibitem[{Chatterjee {et~al.}(2013)Chatterjee, Umbreit, Fregeau, \&
  Rasio}]{Chatterjee_2013b}
Chatterjee, S., Umbreit, S., Fregeau, J.~M., \& Rasio, F.~A. 2013, \mnras, 429,
  2881.
\newblock \url{http://dx.doi.org/10.1093/mnras/sts464}

\bibitem[{{Chomiuk} {et~al.}(2013){Chomiuk}, {Strader}, {Maccarone},
  {Miller-Jones}, {Heinke}, {Noyola}, {Seth}, \& {Ransom}}]{Chomiuk_2013}
{Chomiuk}, L., {Strader}, J., {Maccarone}, T.~J., {et~al.} 2013, \apj, 777, 69,
  \dodoi{10.1088/0004-637X/777/1/69}

\bibitem[{{Clark}(1975)}]{Clark_1975}
{Clark}, G.~W. 1975, \apjl, 199, L143, \dodoi{10.1086/181869}

\bibitem[{{Cool} {et~al.}(1996){Cool}, {Piotto}, \& {King}}]{Cool_1996}
{Cool}, A.~M., {Piotto}, G., \& {King}, I.~R. 1996, \apj, 468, 655,
  \dodoi{10.1086/177723}

\bibitem[{{Dai} {et~al.}(2015){Dai}, {Smith}, {Lin}, {Yue}, {Hobbs}, \&
  {Xu}}]{Dai_2018}
{Dai}, S., {Smith}, M.~C., {Lin}, M.~X., {et~al.} 2015, \apj, 802, 120,
  \dodoi{10.1088/0004-637X/802/2/120}

\bibitem[{{de Rujula} {et~al.}(1991){de Rujula}, {Jetzer}, \&
  {Masso}}]{deRujula_1991}
{de Rujula}, A., {Jetzer}, P., \& {Masso}, E. 1991, \mnras, 250, 348,
  \dodoi{10.1093/mnras/250.2.348}

\bibitem[{{Dotter} {et~al.}(2010){Dotter}, {Sarajedini}, {Anderson},
  {Aparicio}, {Bedin}, {Chaboyer}, {Majewski}, {Mar{\'\i}n-Franch}, {Milone},
  {Paust}, {Piotto}, {Reid}, {Rosenberg}, \& {Siegel}}]{Dotter+2010}
{Dotter}, A., {Sarajedini}, A., {Anderson}, J., {et~al.} 2010, \apj, 708, 698,
  \dodoi{10.1088/0004-637X/708/1/698}

\bibitem[{{Drlica-Wagner} {et~al.}(2019){Drlica-Wagner}, {Mao}, {Adhikari},
  {Armstrong}, {Banerjee}, {Banik}, {Bechtol}, {Bird}, {Boddy}, {Bonaca},
  {Bovy}, {Buckley}, {Bulbul}, {Chang}, {Chapline}, {Cohen-Tanugi}, {Cuoco},
  {Cyr-Racine}, {Dawson}, {D{\'\i}az Rivero}, {Dvorkin}, {Erkal}, {Fassnacht},
  {Garc{\'\i}a-Bellido}, {Giannotti}, {Gluscevic}, {Golovich}, {Hendel},
  {Hezaveh}, {Horiuchi}, {Jee}, {Kaplinghat}, {Keeton}, {Koposov}, {Lam}, {Li},
  {Lu}, {Mandelbaum}, {McDermott}, {McNanna}, {Medford}, {Meyer}, {Marc},
  {Murgia}, {Nadler}, {Necib}, {Nuss}, {Pace}, {Peter}, {Polin},
  {Prescod-Weinstein}, {Read}, {Rosenfeld}, {Shipp}, {Simon}, {Slatyer},
  {Straniero}, {Strigari}, {Tollerud}, {Tyson}, {Wang}, {Wechsler}, {Wittman},
  {Yu}, {Zaharijas}, {Ali-Ha{\"\i}moud}, {Annis}, {Birrer}, {Biswas}, {Blazek},
  {Brooks}, {Buckley-Geer}, {Caputo}, {Charles}, {Digel}, {Dodelson},
  {Flaugher}, {Frieman}, {Gawiser}, {Hearin}, {Hlo{\v{z}}ek}, {Jain},
  {Jeltema}, {Koushiappas}, {Lisanti}, {LoVerde}, {Mishra-Sharma}, {Newman},
  {Nord}, {Nourbakhsh}, {Ritz}, {Robertson}, {S{\'a}nchez-Conde}, {Slosar},
  {Tait}, {Verma}, {Vilalta}, {Walter}, {Yanny}, \& {Zentner}}]{Drlica_2019}
{Drlica-Wagner}, A., {Mao}, Y.-Y., {Adhikari}, S., {et~al.} 2019, arXiv
  e-prints, arXiv:1902.01055.
\newblock \doarXiv{1902.01055}

\bibitem[{{Duquennoy} \& {Mayor}(1991)}]{Duquennoy_1991}
{Duquennoy}, A., \& {Mayor}, M. 1991, \aap, 500, 337

\bibitem[{{Eisenhauer} {et~al.}(2003){Eisenhauer}, {Sch{\"o}del}, {Genzel},
  {Ott}, {Tecza}, {Abuter}, {Eckart}, \& {Alexander}}]{Eisenhauer_2003}
{Eisenhauer}, F., {Sch{\"o}del}, R., {Genzel}, R., {et~al.} 2003, \apjl, 597,
  L121, \dodoi{10.1086/380188}

\bibitem[{{Elson} {et~al.}(1987){Elson}, {Fall}, \& {Freeman}}]{Elson_1987}
{Elson}, R. A.~W., {Fall}, S.~M., \& {Freeman}, K.~C. 1987, \apj, 323, 54,
  \dodoi{10.1086/165807}

\bibitem[{Fragione \& Kocsis(2018)}]{Fragione_2018}
Fragione, G., \& Kocsis, B. 2018, Phys. Rev. Lett., 121, 161103,
  \dodoi{10.1103/PhysRevLett.121.161103}

\bibitem[{Fragione {et~al.}(2018)Fragione, Pavlík, \&
  Banerjee}]{Fragione_2018c}
Fragione, G., Pavlík, V., \& Banerjee, S. 2018, \mnras, 480, 4955,
  \dodoi{10.1093/mnras/sty2234}

\bibitem[{Fregeau {et~al.}(2004)Fregeau, Cheung, Portegies~Zwart, \&
  Rasio}]{fregeau2004stellar}
Fregeau, J.~M., Cheung, P., Portegies~Zwart, S., \& Rasio, F. 2004, \mnras,
  352, 1

\bibitem[{{Fregeau} {et~al.}(2003){Fregeau}, {G{\"u}rkan}, {Joshi}, \&
  {Rasio}}]{Fregeau_2003}
{Fregeau}, J.~M., {G{\"u}rkan}, M.~A., {Joshi}, K.~J., \& {Rasio}, F.~A. 2003,
  \apj, 593, 772, \dodoi{10.1086/376593}

\bibitem[{Fregeau {et~al.}(2002)Fregeau, Joshi, Zwart, \& Rasio}]{Fregeau_2002}
Fregeau, J.~M., Joshi, K.~J., Zwart, S. F.~P., \& Rasio, F.~A. 2002, \apj, 570,
  171, \dodoi{10.1086/339576}

\bibitem[{Fregeau \& Rasio(2007)}]{Fregeau_2007}
Fregeau, J.~M., \& Rasio, F.~A. 2007, \apj, 658, 1047, \dodoi{10.1086/511809}

\bibitem[{{Gaudi}(2012)}]{Gaudi_2012}
{Gaudi}, B.~S. 2012, \araa, 50, 411,
  \dodoi{10.1146/annurev-astro-081811-125518}

\bibitem[{Giersz \& Heggie(2011)}]{Giersz_2011}
Giersz, M., \& Heggie, D.~C. 2011, \mnras, 410, 2698,
  \dodoi{10.1111/j.1365-2966.2010.17648.x}

\bibitem[{{Giersz} {et~al.}(2008){Giersz}, {Heggie}, \& {Hurley}}]{Giersz_2008}
{Giersz}, M., {Heggie}, D.~C., \& {Hurley}, J.~R. 2008, \mnras, 388, 429,
  \dodoi{10.1111/j.1365-2966.2008.13407.x}

\bibitem[{{Giesers} {et~al.}(2018){Giesers}, {Dreizler}, {Husser}, {Kamann},
  {Anglada Escud{\'e}}, {Brinchmann}, {Carollo}, {Roth}, {Weilbacher}, \&
  {Wisotzki}}]{Giesers_2018}
{Giesers}, B., {Dreizler}, S., {Husser}, T.-O., {et~al.} 2018, \mnras, 475,
  L15, \dodoi{10.1093/mnrasl/slx203}

\bibitem[{{Giesers} {et~al.}(2019){Giesers}, {Kamann}, {Dreizler}, {Husser},
  {Askar}, {G{\"o}ttgens}, {Brinchmann}, {Latour}, {Weilbacher}, {Wendt}, \&
  {Roth}}]{Giesers_2019}
{Giesers}, B., {Kamann}, S., {Dreizler}, S., {et~al.} 2019, \aap, 632, A3,
  \dodoi{10.1051/0004-6361/201936203}

\bibitem[{Giesler {et~al.}(2018)Giesler, Clausen, \& Ott}]{Giesler_2018}
Giesler, M., Clausen, D., \& Ott, C.~D. 2018, \mnras, 477, 1853

\bibitem[{{Gillessen} {et~al.}(2009){Gillessen}, {Eisenhauer}, {Trippe},
  {Alexander}, {Genzel}, {Martins}, \& {Ott}}]{Gillessen_2009}
{Gillessen}, S., {Eisenhauer}, F., {Trippe}, S., {et~al.} 2009, \apj, 692,
  1075, \dodoi{10.1088/0004-637X/692/2/1075}

\bibitem[{{Gilliland} {et~al.}(2000){Gilliland}, {Brown}, {Guhathakurta},
  {Sarajedini}, {Milone}, {Albrow}, {Baliber}, {Bruntt}, {Burrows},
  {Charbonneau}, {Choi}, {Cochran}, {Edmonds}, {Frandsen}, {Howell}, {Lin},
  {Marcy}, {Mayor}, {Naef}, {Sigurdsson}, {Stagg}, {Vandenberg}, {Vogt}, \&
  {Williams}}]{Gilliland_2000}
{Gilliland}, R.~L., {Brown}, T.~M., {Guhathakurta}, P., {et~al.} 2000, \apjl,
  545, L47, \dodoi{10.1086/317334}

\bibitem[{{Gould}(2000)}]{Gould_2000}
{Gould}, A. 2000, \apj, 535, 928, \dodoi{10.1086/308865}

\bibitem[{{Griest} {et~al.}(1991){Griest}, {Alcock}, {Axelrod}, {Bennett},
  {Cook}, {Freeman}, {Park}, {Perlmutter}, {Peterson}, {Quinn}, {Rodgers},
  {Stubbs}, \& {MACHO Collaboration}}]{Griest_1991}
{Griest}, K., {Alcock}, C., {Axelrod}, T.~S., {et~al.} 1991, \apjl, 372, L79,
  \dodoi{10.1086/186028}

\bibitem[{Hansen {et~al.}(2002)Hansen, Brewer, Fahlman, Gibson, Ibata, Limongi,
  Rich, Richer, Shara, \& Stetson}]{Hansen_2002}
Hansen, B. M.~S., Brewer, J., Fahlman, G.~G., {et~al.} 2002, \apjl, 574, L155,
  \dodoi{10.1086/342528}

\bibitem[{Hansen {et~al.}(2007)Hansen, Anderson, Brewer, Dotter, Fahlman,
  Hurley, Kalirai, King, Reitzel, Richer, Rich, Shara, \&
  Stetson}]{Hansen_2007}
Hansen, B. M.~S., Anderson, J., Brewer, J., {et~al.} 2007, \apj, 671, 380,
  \dodoi{10.1086/522567}

\bibitem[{{Hansen} {et~al.}(2013{\natexlab{a}}){Hansen}, {Kalirai}, {Anderson},
  {Dotter}, {Richer}, {Rich}, {Shara}, {Fahlman}, {Hurley}, {King}, {Reitzel},
  \& {Stetson}}]{Hansen_2013}
{Hansen}, B.~M.~S., {Kalirai}, J.~S., {Anderson}, J., {et~al.}
  2013{\natexlab{a}}, \nat, 500, 51, \dodoi{10.1038/nature12334}

\bibitem[{{Hansen} {et~al.}(2013{\natexlab{b}}){Hansen}, {Kalirai}, {Anderson},
  {Dotter}, {Richer}, {Rich}, {Shara}, {Fahlman}, {Hurley}, {King}, {Reitzel},
  \& {Stetson}}]{Hansen+2013}
---. 2013{\natexlab{b}}, \nat, 500, 51, \dodoi{10.1038/nature12334}

\bibitem[{Harding {et~al.}(2017)Harding, Stefano, Lépine, Urama, Pham, \&
  Baker}]{Harding_2017}
Harding, A.~J., Stefano, R.~D., Lépine, S., {et~al.} 2017, \mnras, 475, 79,
  \dodoi{10.1093/mnras/stx2985}

\bibitem[{Harris {et~al.}(2020)Harris, Millman, van~der Walt, Gommers,
  Virtanen, Cournapeau, Wieser, Taylor, Berg, Smith, Kern, Picus, Hoyer, van
  Kerkwijk, Brett, Haldane, Fernández~del Río, Wiebe, Peterson,
  Gérard-Marchant, Sheppard, Reddy, Weckesser, Abbasi, Gohlke, \&
  Oliphant}]{2020NumPy-Array}
Harris, C.~R., Millman, K.~J., van~der Walt, S.~J., {et~al.} 2020, Nature, 585,
  357–362, \dodoi{10.1038/s41586-020-2649-2}

\bibitem[{{Harris}(1996)}]{Harris_1996}
{Harris}, W.~E. 1996, \aj, 112, 1487, \dodoi{10.1086/118116}

\bibitem[{{Harris}(2010)}]{Harris_2010}
---. 2010, arXiv e-prints, arXiv:1012.3224.
\newblock \doarXiv{1012.3224}

\bibitem[{{Heggie} \& {Hut}(2003)}]{Heggie_2003}
{Heggie}, D., \& {Hut}, P. 2003, {The Gravitational Million-Body Problem: A
  Multidisciplinary Approach to Star Cluster Dynamics}

\bibitem[{Heinke {et~al.}(2005)Heinke, Grindlay, Edmonds, Cohn, Lugger, Camilo,
  Bogdanov, \& Freire}]{Heinke_2005}
Heinke, C.~O., Grindlay, J.~E., Edmonds, P.~D., {et~al.} 2005, \apj, 625, 796,
  \dodoi{10.1086/429899}

\bibitem[{{H{\'e}nault-Brunet} {et~al.}(2020){H{\'e}nault-Brunet}, {Gieles},
  {Strader}, {Peuten}, {Balbinot}, \& {Douglas}}]{Henault-Brunet_2020}
{H{\'e}nault-Brunet}, V., {Gieles}, M., {Strader}, J., {et~al.} 2020, \mnras,
  491, 113, \dodoi{10.1093/mnras/stz2995}

\bibitem[{H{\'e}non(1971{\natexlab{a}})}]{henon1971monte}
H{\'e}non, M. 1971{\natexlab{a}}, in International Astronomical Union
  Colloquium, Vol.~10, Cambridge University Press, 151--167

\bibitem[{H{\'e}non(1971{\natexlab{b}})}]{henon1971montecluster}
H{\'e}non, M. 1971{\natexlab{b}}, \apss, 13, 284

\bibitem[{Hilditch {et~al.}(2005)Hilditch, Howarth, \& Harries}]{Hilditch_2005}
Hilditch, R.~W., Howarth, I.~D., \& Harries, T.~J. 2005, \mnras, 357, 304,
  \dodoi{10.1111/j.1365-2966.2005.08653.x}

\bibitem[{Hong {et~al.}(2018)Hong, Vesperini, Askar, Giersz, Szkudlarek, \&
  Bulik}]{Hong_2018}
Hong, J., Vesperini, E., Askar, A., {et~al.} 2018, \mnras, 480, 5645,
  \dodoi{10.1093/mnras/sty2211}

\bibitem[{Hunter(2007)}]{Hunter2007}
Hunter, J.~D. 2007, Computing in Science \& Engineering, 9, 90,
  \dodoi{10.1109/MCSE.2007.55}

\bibitem[{Hurley {et~al.}(2000)Hurley, Pols, \& Tout}]{Hurley_2000}
Hurley, J.~R., Pols, O.~R., \& Tout, C.~A. 2000, \mnras, 315, 543,
  \dodoi{10.1046/j.1365-8711.2000.03426.x}

\bibitem[{Hurley {et~al.}(2002)Hurley, Tout, \& Pols}]{Hurley_2002}
Hurley, J.~R., Tout, C.~A., \& Pols, O.~R. 2002, \mnras, 329, 897,
  \dodoi{10.1046/j.1365-8711.2002.05038.x}

\bibitem[{{Ivanova}(2013)}]{Ivanova_2013}
{Ivanova}, N. 2013, \memsai, 84, 123.
\newblock \doarXiv{1301.2203}

\bibitem[{{Ivanova} {et~al.}(2008){Ivanova}, {Heinke}, {Rasio}, {Belczynski},
  \& {Fregeau}}]{Ivanova_2008}
{Ivanova}, N., {Heinke}, C.~O., {Rasio}, F.~A., {Belczynski}, K., \& {Fregeau},
  J.~M. 2008, \mnras, 386, 553, \dodoi{10.1111/j.1365-2966.2008.13064.x}

\bibitem[{{Jetzer} {et~al.}(1998){Jetzer}, {Straessle}, \&
  {Wandeler}}]{Jetzer_1998}
{Jetzer}, P., {Straessle}, M., \& {Wandeler}, U. 1998, \aap, 336, 411.
\newblock \doarXiv{astro-ph/9807101}

\bibitem[{{Joshi} {et~al.}(2001){Joshi}, {Nave}, \& {Rasio}}]{Joshi_2001}
{Joshi}, K.~J., {Nave}, C.~P., \& {Rasio}, F.~A. 2001, \apj, 550, 691,
  \dodoi{10.1086/319771}

\bibitem[{{Joshi} {et~al.}(2000){Joshi}, {Rasio}, \& {Portegies
  Zwart}}]{Joshi_2000}
{Joshi}, K.~J., {Rasio}, F.~A., \& {Portegies Zwart}, S. 2000, \apj, 540, 969,
  \dodoi{10.1086/309350}

\bibitem[{Kains {et~al.}(2016)Kains, Bramich, Sahu, \& Calamida}]{Kains_2016}
Kains, N., Bramich, D.~M., Sahu, K.~C., \& Calamida, A. 2016, \mnras, 460,
  2025, \dodoi{10.1093/mnras/stw1137}

\bibitem[{{Kains} {et~al.}(2018){Kains}, {Calamida}, {Sahu}, {Anderson},
  {Casertano}, \& {Bramich}}]{Kains_2018}
{Kains}, N., {Calamida}, A., {Sahu}, K.~C., {et~al.} 2018, \apj, 867, 37,
  \dodoi{10.3847/1538-4357/aae311}

\bibitem[{{Kalirai} {et~al.}(2012){Kalirai}, {Richer}, {Anderson}, {Dotter},
  {Fahlman}, {Hansen}, {Hurley}, {King}, {Reitzel}, {Rich}, {Shara}, {Stetson},
  \& {Woodley}}]{Kaliari_2012}
{Kalirai}, J.~S., {Richer}, H.~B., {Anderson}, J., {et~al.} 2012, \aj, 143, 11,
  \dodoi{10.1088/0004-6256/143/1/11}

\bibitem[{{Katz}(1975)}]{Katz_1975}
{Katz}, J.~I. 1975, \nat, 253, 698, \dodoi{10.1038/253698a0}

\bibitem[{{King}(1962)}]{King_1962}
{King}, I. 1962, \aj, 67, 471, \dodoi{10.1086/108756}

\bibitem[{{Kirsten} {et~al.}(2021){Kirsten}, {Marcote}, {Nimmo}, {Hessels},
  {Bhardwaj}, {Tendulkar}, {Keimpema}, {Yang}, {Snelders}, {Scholz},
  {Pearlman}, {Law}, {Peters}, {Giroletti}, {Paragi}, {Bassa}, {Hewitt},
  {Bach}, {Bezrukovs}, {Burgay}, {Buttaccio}, {Conway}, {Corongiu}, {Feiler},
  {Forss{\'e}n}, {Gawro{\'n}ski}, {Karuppusamy}, {Kharinov}, {Lindqvist},
  {Maccaferri}, {Melnikov}, {Ould-Boukattine}, {Possenti}, {Surcis}, {Wang},
  {Yuan}, {Aggarwal}, {Anna-Thomas}, {Bower}, {Blaauw}, {Burke-Spolaor},
  {Cassanelli}, {Clarke}, {Fonseca}, {Gaensler}, {Gopinath}, {Kaspi}, {Kassim},
  {Lazio}, {Leung}, {Li}, {Lin}, {Masui}, {Mckinven}, {Michilli}, {Mikhailov},
  {Ng}, {Orbidans}, {Pen}, {Petroff}, {Rahman}, {Ransom}, {Shin}, {Smith},
  {Stairs}, \& {Vlemmings}}]{Kirsten_2021}
{Kirsten}, F., {Marcote}, B., {Nimmo}, K., {et~al.} 2021, arXiv e-prints,
  arXiv:2105.11445.
\newblock \doarXiv{2105.11445}

\bibitem[{{K{\i}z{\i}ltan} {et~al.}(2017){K{\i}z{\i}ltan}, {Baumgardt}, \&
  {Loeb}}]{Kiziltan_2017}
{K{\i}z{\i}ltan}, B., {Baumgardt}, H., \& {Loeb}, A. 2017, \nat, 542, 203,
  \dodoi{10.1038/nature21361}

\bibitem[{{Knigge}(2012)}]{Knigge_2012}
{Knigge}, C. 2012, \memsai, 83, 549.
\newblock \doarXiv{1112.1074}

\bibitem[{Kremer {et~al.}(2018)Kremer, Chatterjee, Rodriguez, \&
  Rasio}]{Kremer_2018a}
Kremer, K., Chatterjee, S., Rodriguez, C.~L., \& Rasio, F.~A. 2018, \apj, 852,
  29, \dodoi{10.3847/1538-4357/aa99df}

\bibitem[{{Kremer} {et~al.}(2019){Kremer}, {Chatterjee}, {Ye}, {Rodriguez}, \&
  {Rasio}}]{Kremer_2019a}
{Kremer}, K., {Chatterjee}, S., {Ye}, C.~S., {Rodriguez}, C.~L., \& {Rasio},
  F.~A. 2019, \apj, 871, 38, \dodoi{10.3847/1538-4357/aaf646}

\bibitem[{Kremer {et~al.}(2021)Kremer, Piro, \& Li}]{Kremer_2021_FRB}
Kremer, K., Piro, A.~L., \& Li, D. 2021, \apjl, 917, L11,
  \dodoi{10.3847/2041-8213/ac13a0}

\bibitem[{{Kremer} {et~al.}(2021){Kremer}, {Rui}, {Weatherford}, {Chatterjee},
  {Fragione}, {Rasio}, {Rodriguez}, \& {Ye}}]{Kremer_2021}
{Kremer}, K., {Rui}, N.~Z., {Weatherford}, N.~C., {et~al.} 2021, \apj, 917, 28,
  \dodoi{10.3847/1538-4357/ac06d4}

\bibitem[{{Kremer} {et~al.}(2018){Kremer}, {Ye}, {Chatterjee}, {Rodriguez}, \&
  {Rasio}}]{Kremer_2018b}
{Kremer}, K., {Ye}, C.~S., {Chatterjee}, S., {Rodriguez}, C.~L., \& {Rasio},
  F.~A. 2018, \apjl, 855, L15, \dodoi{10.3847/2041-8213/aab26c}

\bibitem[{Kremer {et~al.}(2019{\natexlab{a}})Kremer, Ye, Chatterjee, Rodriguez,
  \& Rasio}]{Kremer_2019b}
Kremer, K., Ye, C.~S., Chatterjee, S., Rodriguez, C.~L., \& Rasio, F.~A.
  2019{\natexlab{a}}, Proceedings of the International Astronomical Union, 14,
  357–366, \dodoi{10.1017/S1743921319007269}

\bibitem[{Kremer {et~al.}(2019{\natexlab{b}})Kremer, Rodriguez, Amaro-Seoane,
  Breivik, Chatterjee, Katz, Larson, Rasio, Samsing, Ye, \&
  Zevin}]{Kremer_2019d}
Kremer, K., Rodriguez, C.~L., Amaro-Seoane, P., {et~al.} 2019{\natexlab{b}},
  Phys. Rev. D, 99, 063003, \dodoi{10.1103/PhysRevD.99.063003}

\bibitem[{Kremer {et~al.}(2020)Kremer, Ye, Rui, Weatherford, Chatterjee,
  Fragione, Rodriguez, Spera, \& Rasio}]{Kremer_2020}
Kremer, K., Ye, C.~S., Rui, N.~Z., {et~al.} 2020, \apjs, 247, 48,
  \dodoi{10.3847/1538-4365/ab7919}

\bibitem[{{Kroupa}(2001)}]{Kroupa_2001}
{Kroupa}, P. 2001, \mnras, 322, 231, \dodoi{10.1046/j.1365-8711.2001.04022.x}

\bibitem[{{Kuranov} \& {Postnov}(2006)}]{Kuranov_2006}
{Kuranov}, A.~G., \& {Postnov}, K.~A. 2006, Astronomy Letters, 32, 393,
  \dodoi{10.1134/S106377370606003X}

\bibitem[{{Lam} {et~al.}(2022){Lam}, {Lu}, {Udalski}, {Bond}, {Bennett},
  {Skowron}, {Mroz}, {Poleski}, {Sumi}, {Szymanski}, {Kozlowski},
  {Pietrukowicz}, {Soszynski}, {Ulaczyk}, {Wyrzykowski}, {Miyazaki}, {Suzuki},
  {Koshimoto}, {Rattenbury}, {Hosek}, {Abe}, {Barry}, {Bhattacharya}, {Fukui},
  {Fujii}, {Hirao}, {Itow}, {Kirikawa}, {Kondo}, {Matsubara}, {Matsumoto},
  {Muraki}, {Olmschenk}, {Ranc}, {Okamura}, {Satoh}, {Ishitani Silva}, {Toda},
  {Tristram}, {Vandorou}, {Yama}, {Abrams}, {Agarwal}, {Rose}, \&
  {Terry}}]{Lam_2022}
{Lam}, C.~Y., {Lu}, J.~R., {Udalski}, A., {et~al.} 2022, arXiv e-prints,
  arXiv:2202.01903.
\newblock \doarXiv{2202.01903}

\bibitem[{{LSST Science Collaboration} {et~al.}(2009){LSST Science
  Collaboration}, {Abell}, {Allison}, {Anderson}, {Andrew}, {Angel}, {Armus},
  {Arnett}, {Asztalos}, {Axelrod}, {Bailey}, {Ballantyne}, {Bankert},
  {Barkhouse}, {Barr}, {Barrientos}, {Barth}, {Bartlett}, {Becker}, {Becla},
  {Beers}, {Bernstein}, {Biswas}, {Blanton}, {Bloom}, {Bochanski}, {Boeshaar},
  {Borne}, {Bradac}, {Brandt}, {Bridge}, {Brown}, {Brunner}, {Bullock},
  {Burgasser}, {Burge}, {Burke}, {Cargile}, {Chandrasekharan}, {Chartas},
  {Chesley}, {Chu}, {Cinabro}, {Claire}, {Claver}, {Clowe}, {Connolly}, {Cook},
  {Cooke}, {Cooray}, {Covey}, {Culliton}, {de Jong}, {de Vries}, {Debattista},
  {Delgado}, {Dell'Antonio}, {Dhital}, {Di Stefano}, {Dickinson}, {Dilday},
  {Djorgovski}, {Dobler}, {Donalek}, {Dubois-Felsmann}, {Durech},
  {Eliasdottir}, {Eracleous}, {Eyer}, {Falco}, {Fan}, {Fassnacht}, {Ferguson},
  {Fernandez}, {Fields}, {Finkbeiner}, {Figueroa}, {Fox}, {Francke}, {Frank},
  {Frieman}, {Fromenteau}, {Furqan}, {Galaz}, {Gal-Yam}, {Garnavich},
  {Gawiser}, {Geary}, {Gee}, {Gibson}, {Gilmore}, {Grace}, {Green}, {Gressler},
  {Grillmair}, {Habib}, {Haggerty}, {Hamuy}, {Harris}, {Hawley}, {Heavens},
  {Hebb}, {Henry}, {Hileman}, {Hilton}, {Hoadley}, {Holberg}, {Holman},
  {Howell}, {Infante}, {Ivezic}, {Jacoby}, {Jain}, {R}, {Jedicke}, {Jee},
  {Garrett Jernigan}, {Jha}, {Johnston}, {Jones}, {Juric}, {Kaasalainen},
  {Styliani}, {Kafka}, {Kahn}, {Kaib}, {Kalirai}, {Kantor}, {Kasliwal},
  {Keeton}, {Kessler}, {Knezevic}, {Kowalski}, {Krabbendam}, {Krughoff},
  {Kulkarni}, {Kuhlman}, {Lacy}, {Lepine}, {Liang}, {Lien}, {Lira}, {Long},
  {Lorenz}, {Lotz}, {Lupton}, {Lutz}, {Macri}, {Mahabal}, {Mandelbaum},
  {Marshall}, {May}, {McGehee}, {Meadows}, {Meert}, {Milani}, {Miller},
  {Miller}, {Mills}, {Minniti}, {Monet}, {Mukadam}, {Nakar}, {Neill}, {Newman},
  {Nikolaev}, {Nordby}, {O'Connor}, {Oguri}, {Oliver}, {Olivier}, {Olsen},
  {Olsen}, {Olszewski}, {Oluseyi}, {Padilla}, {Parker}, {Pepper}, {Peterson},
  {Petry}, {Pinto}, {Pizagno}, {Popescu}, {Prsa}, {Radcka}, {Raddick},
  {Rasmussen}, {Rau}, {Rho}, {Rhoads}, {Richards}, {Ridgway}, {Robertson},
  {Roskar}, {Saha}, {Sarajedini}, {Scannapieco}, {Schalk}, {Schindler},
  {Schmidt}, {Schmidt}, {Schneider}, {Schumacher}, {Scranton}, {Sebag},
  {Seppala}, {Shemmer}, {Simon}, {Sivertz}, {Smith}, {Allyn Smith}, {Smith},
  {Spitz}, {Stanford}, {Stassun}, {Strader}, {Strauss}, {Stubbs}, {Sweeney},
  {Szalay}, {Szkody}, {Takada}, {Thorman}, {Trilling}, {Trimble}, {Tyson}, {Van
  Berg}, {Vanden Berk}, {VanderPlas}, {Verde}, {Vrsnak}, {Walkowicz},
  {Wandelt}, {Wang}, {Wang}, {Warner}, {Wechsler}, {West}, {Wiecha},
  {Williams}, {Willman}, {Wittman}, {Wolff}, {Wood-Vasey}, {Wozniak}, {Young},
  {Zentner}, \& {Zhan}}]{LSST_2009}
{LSST Science Collaboration}, {Abell}, P.~A., {Allison}, J., {et~al.} 2009,
  arXiv e-prints, arXiv:0912.0201.
\newblock \doarXiv{0912.0201}

\bibitem[{Lu {et~al.}(2016)Lu, Sinukoff, Ofek, Udalski, \& Kozlowski}]{Lu_2016}
Lu, J.~R., Sinukoff, E., Ofek, E.~O., Udalski, A., \& Kozlowski, S. 2016, \apj,
  830, 41, \dodoi{10.3847/0004-637x/830/1/41}

\bibitem[{{Lyne} {et~al.}(1987){Lyne}, {Brinklow}, {Middleditch}, {Kulkarni},
  \& {Backer}}]{Lyne_1987}
{Lyne}, A.~G., {Brinklow}, A., {Middleditch}, J., {Kulkarni}, S.~R., \&
  {Backer}, D.~C. 1987, \nat, 328, 399, \dodoi{10.1038/328399a0}

\bibitem[{Maccarone {et~al.}(2007)Maccarone, Kundu, Zepf, \&
  Rhode}]{Maccarone_2007}
Maccarone, T.~J., Kundu, A., Zepf, S.~E., \& Rhode, K.~L. 2007, Nature, 445,
  183, \dodoi{10.1038/nature05434}

\bibitem[{Mackey {et~al.}(2007)Mackey, Wilkinson, Davies, \&
  Gilmore}]{Mackey_2007}
Mackey, A.~D., Wilkinson, M.~I., Davies, M.~B., \& Gilmore, G.~F. 2007, \mnras,
  379, L40, \dodoi{10.1111/j.1745-3933.2007.00330.x}

\bibitem[{Mackey {et~al.}(2008)Mackey, Wilkinson, Davies, \&
  Gilmore}]{Mackey_2008}
---. 2008, \mnras, 386, 65, \dodoi{10.1111/j.1365-2966.2008.13052.x}

\bibitem[{{McGill} {et~al.}(2018){McGill}, {Smith}, {Evans}, {Belokurov}, \&
  {Smart}}]{McGill_2018}
{McGill}, P., {Smith}, L.~C., {Evans}, N.~W., {Belokurov}, V., \& {Smart},
  R.~L. 2018, \mnras, 478, L29, \dodoi{10.1093/mnrasl/sly066}

\bibitem[{{McGill} {et~al.}(2019){McGill}, {Smith}, {Evans}, {Belokurov}, \&
  {Zhang}}]{McGill_2019}
{McGill}, P., {Smith}, L.~C., {Evans}, N.~W., {Belokurov}, V., \& {Zhang},
  Z.~H. 2019, \mnras, 483, 4210, \dodoi{10.1093/mnras/sty3344}

\bibitem[{{M}c{K}inney(2010)}]{mckinney-proc-scipy-2010}
{M}c{K}inney, W. 2010, in {P}roceedings of the 9th {P}ython in {S}cience
  {C}onference, ed. {S}t\'efan van~der {W}alt \& {J}arrod {M}illman, 56 -- 61,
  \dodoi{10.25080/Majora-92bf1922-00a}

\bibitem[{Merritt {et~al.}(2004)Merritt, Piatek, Zwart, \&
  Hemsendorf}]{Merritt_2004}
Merritt, D., Piatek, S., Zwart, S.~P., \& Hemsendorf, M. 2004, \apjl, 608, L25,
  \dodoi{10.1086/422252}

\bibitem[{Miller-Jones {et~al.}(2015)Miller-Jones, Strader, Heinke, Maccarone,
  van~den Berg, Knigge, Chomiuk, Noyola, Russell, Seth, \&
  Sivakoff}]{Miller-Jones_2015}
Miller-Jones, J. C.~A., Strader, J., Heinke, C.~O., {et~al.} 2015, \mnras, 453,
  3918, \dodoi{10.1093/mnras/stv1869}

\bibitem[{Monaco {et~al.}(2004)Monaco, Pancino, Ferraro, \&
  Bellazzini}]{Monaco_2004}
Monaco, L., Pancino, E., Ferraro, F.~R., \& Bellazzini, M. 2004, \mnras, 349,
  1278, \dodoi{10.1111/j.1365-2966.2004.07599.x}

\bibitem[{Moody \& Sigurdsson(2008)}]{Moody_2008}
Moody, K., \& Sigurdsson, S. 2008, \apj, 690, 1370,
  \dodoi{10.1088/0004-637x/690/2/1370}

\bibitem[{Morscher {et~al.}(2015)Morscher, Pattabiraman, Rodriguez, Rasio, \&
  Umbreit}]{Morscher_2015}
Morscher, M., Pattabiraman, B., Rodriguez, C., Rasio, F.~A., \& Umbreit, S.
  2015, \apj, 800, 9, \dodoi{10.1088/0004-637x/800/1/9}

\bibitem[{{Mroz} {et~al.}(2021){Mroz}, {Udalski}, {Wyrzykowski}, {Skowron},
  {Poleski}, {Szymanski}, {Soszynski}, \& {Ulaczyk}}]{Mroz_2021}
{Mroz}, P., {Udalski}, A., {Wyrzykowski}, L., {et~al.} 2021, arXiv e-prints,
  arXiv:2107.13697.
\newblock \doarXiv{2107.13697}

\bibitem[{{Ofek}(2018)}]{Ofek_2018}
{Ofek}, E.~O. 2018, \apj, 866, 144, \dodoi{10.3847/1538-4357/aadfeb}

\bibitem[{{Paczynski}(1986)}]{Paczynski_1986}
{Paczynski}, B. 1986, \apj, 304, 1, \dodoi{10.1086/164140}

\bibitem[{{Paczynski}(1991)}]{Paczynski_1991}
---. 1991, \apjl, 371, L63, \dodoi{10.1086/186003}

\bibitem[{{Paczynski}(1994)}]{Paczynski_1994}
---. 1994, \actaa, 44, 235.
\newblock \doarXiv{astro-ph/9407096}

\bibitem[{{Paczy\'{n}ski}(1996)}]{Paczynski_1996}
{Paczy\'{n}ski}, B. 1996, Annual Review of Astronomy and Astrophysics, 34, 419,
  \dodoi{10.1146/annurev.astro.34.1.419}

\bibitem[{{Pattabiraman} {et~al.}(2013){Pattabiraman}, {Umbreit}, {Liao},
  {Choudhary}, {Kalogera}, {Memik}, \& {Rasio}}]{Pattabiraman_2013}
{Pattabiraman}, B., {Umbreit}, S., {Liao}, W.-k., {et~al.} 2013, \apjs, 204,
  15, \dodoi{10.1088/0067-0049/204/2/15}

\bibitem[{Peuten {et~al.}(2016)Peuten, Zocchi, Gieles, Gualandris, \&
  Hénault-Brunet}]{Peuten_2016}
Peuten, M., Zocchi, A., Gieles, M., Gualandris, A., \& Hénault-Brunet, V.
  2016, \mnras, 462, 2333, \dodoi{10.1093/mnras/stw1726}

\bibitem[{{Pietrukowicz} {et~al.}(2012){Pietrukowicz}, {Minniti}, {Jetzer},
  {Alonso-Garc{\'\i}a}, \& {Udalski}}]{Pietrukowicz_2012}
{Pietrukowicz}, P., {Minniti}, D., {Jetzer}, P., {Alonso-Garc{\'\i}a}, J., \&
  {Udalski}, A. 2012, \apjl, 744, L18, \dodoi{10.1088/2041-8205/744/2/L18}

\bibitem[{{Ransom}(2008)}]{Ransom_2008}
{Ransom}, S.~M. 2008, in Dynamical Evolution of Dense Stellar Systems, ed.
  E.~{Vesperini}, M.~{Giersz}, \& A.~{Sills}, Vol. 246, 291--300,
  \dodoi{10.1017/S1743921308015810}

\bibitem[{Reback {et~al.}(2021)Reback, McKinney, jbrockmendel, den Bossche,
  Augspurger, Cloud, Hawkins, gfyoung, Sinhrks, Roeschke, Klein, Petersen,
  Tratner, She, Ayd, Naveh, patrick, Garcia, Schendel, Hayden, Saxton,
  Jancauskas, Gorelli, Shadrach, McMaster, Battiston, Seabold, Dong, chris b1,
  \& h~vetinari}]{jeff_reback_2021_4681666}
Reback, J., McKinney, W., jbrockmendel, {et~al.} 2021, pandas-dev/pandas:
  Pandas 1.2.4, v1.2.4,  Zenodo, \dodoi{10.5281/zenodo.4681666}

\bibitem[{Renzini {et~al.}(1996)Renzini, Bragaglia, Ferraro, Gilmozzi,
  Ortolani, Holberg, Liebert, Wesemael, \& Bohlin}]{Renzini_1996}
Renzini, A., Bragaglia, A., Ferraro, F.~R., {et~al.} 1996, \apjl, 465, L23,
  \dodoi{10.1086/310128}

\bibitem[{{Richer} {et~al.}(1995){Richer}, {Fahlman}, {Ibata}, {Stetson},
  {Bell}, {Bolte}, {Bond}, {Harris}, {Hesser}, {Mandushev}, {Pryor}, \&
  {Vandenberg}}]{Richer_1995}
{Richer}, H.~B., {Fahlman}, G.~G., {Ibata}, R.~A., {et~al.} 1995, \apjl, 451,
  L17, \dodoi{10.1086/309674}

\bibitem[{Richer {et~al.}(1997)Richer, Fahlman, Ibata, Pryor, Bell, Bolte,
  Bond, Harris, Hesser, Holland, Ivanans, Mandushev, Stetson, \&
  Wood}]{Richer_1997}
Richer, H.~B., Fahlman, G.~G., Ibata, R.~A., {et~al.} 1997, \apj, 484, 741,
  \dodoi{10.1086/304379}

\bibitem[{{Rodriguez} {et~al.}(2018){Rodriguez}, {Amaro-Seoane}, {Chatterjee},
  \& {Rasio}}]{Rodriguez_2018}
{Rodriguez}, C.~L., {Amaro-Seoane}, P., {Chatterjee}, S., \& {Rasio}, F.~A.
  2018, \prl, 120, 151101, \dodoi{10.1103/PhysRevLett.120.151101}

\bibitem[{Rodriguez {et~al.}(2016)Rodriguez, Chatterjee, \&
  Rasio}]{Rodriguez_2016}
Rodriguez, C.~L., Chatterjee, S., \& Rasio, F.~A. 2016, Phys. Rev. D, 93,
  084029, \dodoi{10.1103/PhysRevD.93.084029}

\bibitem[{Rodriguez {et~al.}(2015)Rodriguez, Morscher, Pattabiraman,
  Chatterjee, Haster, \& Rasio}]{Rodriguez_2015}
Rodriguez, C.~L., Morscher, M., Pattabiraman, B., {et~al.} 2015, Phys. Rev.
  Lett., 115, 051101, \dodoi{10.1103/PhysRevLett.115.051101}

\bibitem[{{Rodriguez} {et~al.}(2016){Rodriguez}, {Morscher}, {Wang},
  {Chatterjee}, {Rasio}, \& {Spurzem}}]{Rodriguez+2016million}
{Rodriguez}, C.~L., {Morscher}, M., {Wang}, L., {et~al.} 2016, \mnras, 463,
  2109, \dodoi{10.1093/mnras/stw2121}

\bibitem[{{Rodriguez} {et~al.}(2021){Rodriguez}, {Weatherford}, {Coughlin}, \&
  et~al.}]{Rodriguez2021}
{Rodriguez}, C.~L., {Weatherford}, N.~C., {Coughlin}, S.~C., \& et~al. 2021,
  arXiv e-prints, arXiv:2106.02643.
\newblock \doarXiv{2106.02643}

\bibitem[{{Rui} {et~al.}(2021{\natexlab{a}}){Rui}, {Kremer}, {Weatherford},
  {Chatterjee}, {Rasio}, {Rodriguez}, \& {Ye}}]{Rui_2021}
{Rui}, N.~Z., {Kremer}, K., {Weatherford}, N.~C., {et~al.} 2021{\natexlab{a}},
  \apj, 912, 102, \dodoi{10.3847/1538-4357/abed49}

\bibitem[{{Rui} {et~al.}(2021{\natexlab{b}}){Rui}, {Weatherford}, {Kremer},
  {Chatterjee}, {Fragione}, {Rasio}, {Rodriguez}, \& {Ye}}]{Rui_2021b}
{Rui}, N.~Z., {Weatherford}, N.~C., {Kremer}, K., {et~al.} 2021{\natexlab{b}},
  Research Notes of the American Astronomical Society, 5, 47,
  \dodoi{10.3847/2515-5172/abee77}

\bibitem[{{Rybicki} {et~al.}(2022){Rybicki}, {Wyrzykowski}, {Bachelet},
  {Cassan}, {Zieli{\'n}ski}, {Gould}, {Calchi Novati}, {Yee}, {Ryu},
  {Gromadzki}, {Miko{\l}ajczyk}, {Ihanec}, {Kruszy{\'n}ska}, {Hambsch},
  {Zo{\l}a}, {Fossey}, {Awiphan}, {Nakharutai}, {Lewis}, {Olivares E.},
  {Hodgkin}, {Delgado}, {Breedt}, {Harrison}, {van Leeuwen}, {Rixon}, {Wevers},
  {Yoldas}, {Udalski}, {Szyma{\'n}ski}, {Soszy{\'n}ski}, {Pietrukowicz},
  {Koz{\l}owski}, {Skowron}, {Poleski}, {Ulaczyk}, {Mr{\'o}z}, {Iwanek},
  {Wrona}, {Street}, {Tsapras}, {Hundertmark}, {Dominik}, {Beichman}, {Bryden},
  {Carey}, {Gaudi}, {Henderson}, {Shvartzvald}, {Zang}, {Zhu}, {Christie},
  {Green}, {Hennerley}, {McCormick}, {Monard}, {Natusch}, {Pogge}, {Gezer},
  {Gurgul}, {Kaczmarek}, {Konacki}, {Lam}, {Maskoliunas}, {Pakstiene},
  {Ratajczak}, {Stankeviciute}, {Zdanavicius}, \&
  {Zi{\'o}{\l}kowska}}]{Rybicki_2022}
{Rybicki}, K.~A., {Wyrzykowski}, {\L}., {Bachelet}, E., {et~al.} 2022, \aap,
  657, A18, \dodoi{10.1051/0004-6361/202039542}

\bibitem[{{Safonova} \& {Rahvar}(2007)}]{Safonova_2007}
{Safonova}, M., \& {Rahvar}, S. 2007, in Black Holes from Stars to Galaxies --
  Across the Range of Masses, ed. V.~{Karas} \& G.~{Matt}, Vol. 238, 439--440,
  \dodoi{10.1017/S1743921307005844}

\bibitem[{{Sahu} {et~al.}(2001){Sahu}, {Casertano}, {Livio}, {Gilliland},
  {Panagia}, {Albrow}, \& {Potter}}]{Sahu_2001}
{Sahu}, K.~C., {Casertano}, S., {Livio}, M., {et~al.} 2001, \nat, 411, 1022

\bibitem[{{Sahu} {et~al.}(2017){Sahu}, {Anderson}, {Casertano}, {Bond},
  {Bergeron}, {Nelan}, {Pueyo}, {Brown}, {Bellini}, {Levay}, {Sokol}, {aff1},
  {Dominik}, {Calamida}, {Kains}, \& {Livio}}]{Sahu_2017}
{Sahu}, K.~C., {Anderson}, J., {Casertano}, S., {et~al.} 2017, Science, 356,
  1046, \dodoi{10.1126/science.aal2879}

\bibitem[{{Sahu} {et~al.}(2022){Sahu}, {Anderson}, {Casertano}, {Bond},
  {Udalski}, {Dominik}, {Calamida}, {Bellini}, {Brown}, {Rejkuba}, {Bajaj},
  {Kains}, {Ferguson}, {Fryer}, {Yock}, {Mroz}, {Kozlowski}, {Pietrukowicz},
  {Poleski}, {Skowron}, {Soszynski}, {Szymanski}, {Ulaczyk}, {Wyrzykowski},
  {Beaulieu}, {Marquette}, {Cole}, {Hill}, {Dieters}, {Coutures},
  {Dominis-Prester}, {Bachelet}, {Menzies}, {Albrow}, {Pollard}, {Gould},
  {Yee}, {Allen}, {de Almeida}, {Christie}, {Drummond}, {Gal-Yam}, {Gorbikov},
  {Jablonski}, {Lee}, {Maoz}, {Manulis}, {McCormick}, {Natusch}, {Pogge},
  {Shvartzvald}, {Jorgensen}, {Alsubai}, {Andersen}, {Bozza}, {Calchi Novati},
  {Hinse}, {Hundertmark}, {Husser}, {Kerins}, {Longa-Pena}, {Mancini}, {Penny},
  {Rahvar}, {Ricci}, {Sajadian}, {Skottfelt}, {Snodgrass}, {Southworth},
  {Tregloan-Reed}, {Wambsganss}, {Wertz}, {Tsapras}, {Street}, {Bramich},
  {Horne}, \& {Steele}}]{Sahu_2022}
---. 2022, arXiv e-prints, arXiv:2201.13296.
\newblock \doarXiv{2201.13296}

\bibitem[{Samsing \& D’Orazio(2018)}]{Samsing_2018}
Samsing, J., \& D’Orazio, D.~J. 2018, \mnras, 481, 5445,
  \dodoi{10.1093/mnras/sty2334}

\bibitem[{{Sarajedini} {et~al.}(2007){Sarajedini}, {Bedin}, {Chaboyer},
  {Dotter}, {Siegel}, {Anderson}, {Aparicio}, {King}, {Majewski},
  {Mar{\'\i}n-Franch}, {Piotto}, {Reid}, \& {Rosenberg}}]{Sarajedini_2007}
{Sarajedini}, A., {Bedin}, L.~R., {Chaboyer}, B., {et~al.} 2007, \aj, 133,
  1658, \dodoi{10.1086/511979}

\bibitem[{{Shishkovsky} {et~al.}(2018){Shishkovsky}, {Strader}, {Chomiuk},
  {Bahramian}, {Tremou}, {Li}, {Salinas}, {Tudor}, {Miller-Jones}, {Maccarone},
  {Heinke}, \& {Sivakoff}}]{Shishkovsky_2018}
{Shishkovsky}, L., {Strader}, J., {Chomiuk}, L., {et~al.} 2018, \apj, 855, 55,
  \dodoi{10.3847/1538-4357/aaadb1}

\bibitem[{{Sigurdsson} \& {Phinney}(1995)}]{Sigurdsson_1995}
{Sigurdsson}, S., \& {Phinney}, E.~S. 1995, \apjs, 99, 609,
  \dodoi{10.1086/192199}

\bibitem[{{Spitzer}(1969)}]{Spitzer_1969}
{Spitzer}, Lyman, J. 1969, \apjl, 158, L139, \dodoi{10.1086/180451}

\bibitem[{{Spitzer}(1987)}]{Spitzer_1987}
{Spitzer}, L. 1987, {Dynamical evolution of globular clusters}

\bibitem[{Stefano(2012)}]{DiStefano_2012}
Stefano, R.~D. 2012, \apjs, 201, 20, \dodoi{10.1088/0067-0049/201/2/20}

\bibitem[{{Strader} {et~al.}(2012{\natexlab{a}}){Strader}, {Chomiuk},
  {Maccarone}, {Miller-Jones}, \& {Seth}}]{Strader_2012}
{Strader}, J., {Chomiuk}, L., {Maccarone}, T.~J., {Miller-Jones}, J. C.~A., \&
  {Seth}, A.~C. 2012{\natexlab{a}}, \nat, 490, 71, \dodoi{10.1038/nature11490}

\bibitem[{{Strader} {et~al.}(2012{\natexlab{b}}){Strader}, {Chomiuk},
  {Maccarone}, {Miller-Jones}, {Seth}, {Heinke}, \& {Sivakoff}}]{Strader_2012b}
{Strader}, J., {Chomiuk}, L., {Maccarone}, T.~J., {et~al.} 2012{\natexlab{b}},
  \apjl, 750, L27, \dodoi{10.1088/2041-8205/750/2/L27}

\bibitem[{{Sumi} {et~al.}(2011){Sumi}, {Kamiya}, {Bennett}, {Bond}, {Abe},
  {Botzler}, {Fukui}, {Furusawa}, {Hearnshaw}, {Itow}, {Kilmartin}, {Korpela},
  {Lin}, {Ling}, {Masuda}, {Matsubara}, {Miyake}, {Motomura}, {Muraki},
  {Nagaya}, {Nakamura}, {Ohnishi}, {Okumura}, {Perrott}, {Rattenbury}, {Saito},
  {Sako}, {Sullivan}, {Sweatman}, {Tristram}, {Udalski}, {Szyma{\'n}ski},
  {Kubiak}, {Pietrzy{\'n}ski}, {Poleski}, {Soszy{\'n}ski}, {Wyrzykowski},
  {Ulaczyk}, \& {Microlensing Observations in Astrophysics (MOA)
  Collaboration}}]{Sumi_2011}
{Sumi}, T., {Kamiya}, K., {Bennett}, D.~P., {et~al.} 2011, \nat, 473, 349,
  \dodoi{10.1038/nature10092}

\bibitem[{{Thompson} {et~al.}(2010){Thompson}, {Kaluzny}, {Rucinski},
  {Krzeminski}, {Pych}, {Dotter}, \& {Burley}}]{Thompson+2010}
{Thompson}, I.~B., {Kaluzny}, J., {Rucinski}, S.~M., {et~al.} 2010, \aj, 139,
  329, \dodoi{10.1088/0004-6256/139/2/329}

\bibitem[{{Thompson} {et~al.}(2020){Thompson}, {Udalski}, {Dotter}, {Rozyczka},
  {Schwarzenberg-Czerny}, {Pych}, {Beletsky}, {Burley}, {Marshall},
  {McWilliam}, {Morrell}, {Osip}, {Monson}, {Persson}, {Szyma{\'n}ski},
  {Soszy{\'n}ski}, {Poleski}, {Ulaczyk}, {Wyrzykowski}, {Koz{\l}owski},
  {Mr{\'o}z}, {Pietrukowicz}, \& {Skowron}}]{Thompson+2020}
{Thompson}, I.~B., {Udalski}, A., {Dotter}, A., {et~al.} 2020, \mnras, 492,
  4254, \dodoi{10.1093/mnras/staa032}

\bibitem[{{Trager} {et~al.}(1995){Trager}, {King}, \&
  {Djorgovski}}]{Trager1995}
{Trager}, S.~C., {King}, I.~R., \& {Djorgovski}, S. 1995, \aj, 109, 218,
  \dodoi{10.1086/117268}

\bibitem[{{Tremou} {et~al.}(2018){Tremou}, {Strader}, {Chomiuk}, {Shishkovsky},
  {Maccarone}, {Miller-Jones}, {Tudor}, {Heinke}, {Sivakoff}, {Seth}, \&
  {Noyola}}]{Tremou_2018}
{Tremou}, E., {Strader}, J., {Chomiuk}, L., {et~al.} 2018, \apj, 862, 16,
  \dodoi{10.3847/1538-4357/aac9b9}

\bibitem[{{Tucholke}(1992)}]{Tucholke_1992}
{Tucholke}, H.~J. 1992, \aaps, 93, 293

\bibitem[{{Udalski}(2003)}]{Udalski_2003}
{Udalski}, A. 2003, \actaa, 53, 291.
\newblock \doarXiv{astro-ph/0401123}

\bibitem[{{Udalski} {et~al.}(1994){Udalski}, {Szymanski}, {Kaluzny}, {Kubiak},
  {Mateo}, {Krzeminski}, \& {Paczynski}}]{Udalski_1994}
{Udalski}, A., {Szymanski}, M., {Kaluzny}, J., {et~al.} 1994, \actaa, 44, 227.
\newblock \doarXiv{astro-ph/9408026}

\bibitem[{{Udalski} {et~al.}(2015){Udalski}, {Szyma{\'n}ski}, \&
  {Szyma{\'n}ski}}]{Udalski_2015}
{Udalski}, A., {Szyma{\'n}ski}, M.~K., \& {Szyma{\'n}ski}, G. 2015, \actaa, 65,
  1.
\newblock \doarXiv{1504.05966}

\bibitem[{Umbreit {et~al.}(2012)Umbreit, Fregeau, Chatterjee, \&
  Rasio}]{Umbreit_2012}
Umbreit, S., Fregeau, J.~M., Chatterjee, S., \& Rasio, F.~A. 2012, \apj, 750,
  31.
\newblock \url{http://dx.doi.org/10.1088/0004-637X/750/1/31}

\bibitem[{{van Saders} \& {Gaudi}(2011)}]{VanSaders_2011}
{van Saders}, J.~L., \& {Gaudi}, B.~S. 2011, \apj, 729, 63,
  \dodoi{10.1088/0004-637X/729/1/63}

\bibitem[{{VandenBerg} {et~al.}(2013){VandenBerg}, {Brogaard}, {Leaman}, \&
  {Casagrande}}]{VandenBerg+2013}
{VandenBerg}, D.~A., {Brogaard}, K., {Leaman}, R., \& {Casagrande}, L. 2013,
  \apj, 775, 134, \dodoi{10.1088/0004-637X/775/2/134}

\bibitem[{{Vanhollebeke} {et~al.}(2009){Vanhollebeke}, {Groenewegen}, \&
  {Girardi}}]{Vanhollebeke_2009}
{Vanhollebeke}, E., {Groenewegen}, M.~A.~T., \& {Girardi}, L. 2009, \aap, 498,
  95, \dodoi{10.1051/0004-6361/20078472}

\bibitem[{Verbunt {et~al.}(1984)Verbunt, van Paradijs, \& Elson}]{Verbunt_1984}
Verbunt, F., van Paradijs, J., \& Elson, R. 1984, \mnras, 210, 899,
  \dodoi{10.1093/mnras/210.4.899}

\bibitem[{{Vitral} \& {Mamon}(2021)}]{Vitral_2021}
{Vitral}, E., \& {Mamon}, G.~A. 2021, \aap, 646, A63,
  \dodoi{10.1051/0004-6361/202039650}

\bibitem[{Weatherford {et~al.}(2020)Weatherford, Chatterjee, Kremer, \&
  Rasio}]{Weatherford_2020}
Weatherford, N.~C., Chatterjee, S., Kremer, K., \& Rasio, F.~A. 2020, \apj,
  898, 162, \dodoi{10.3847/1538-4357/ab9f98}

\bibitem[{Weatherford {et~al.}(2018)Weatherford, Chatterjee, Rodriguez, \&
  Rasio}]{Weatherford_2018}
Weatherford, N.~C., Chatterjee, S., Rodriguez, C.~L., \& Rasio, F.~A. 2018,
  \apj, 864, 13, \dodoi{10.3847/1538-4357/aad63d}

\bibitem[{{Weldrake} {et~al.}(2008){Weldrake}, {Sackett}, \&
  {Bridges}}]{Weldrake_2008}
{Weldrake}, D. T.~F., {Sackett}, P.~D., \& {Bridges}, T.~J. 2008, \apj, 674,
  1117, \dodoi{10.1086/524917}

\bibitem[{{Wyrzykowski} \& {Mandel}(2020)}]{Wyrzykowski_2020}
{Wyrzykowski}, {\L}., \& {Mandel}, I. 2020, \aap, 636, A20,
  \dodoi{10.1051/0004-6361/201935842}

\bibitem[{{Wyrzykowski} {et~al.}(2016){Wyrzykowski}, {Kostrzewa-Rutkowska},
  {Skowron}, {Rybicki}, {Mr{\'o}z}, {Koz{\l}owski}, {Udalski}, {Szyma{\'n}ski},
  {Pietrzy{\'n}ski}, {Soszy{\'n}ski}, {Ulaczyk}, {Pietrukowicz}, {Poleski},
  {Pawlak}, {I{\l}kiewicz}, \& {Rattenbury}}]{Wyrzykowski_2016}
{Wyrzykowski}, {\L}., {Kostrzewa-Rutkowska}, Z., {Skowron}, J., {et~al.} 2016,
  \mnras, 458, 3012, \dodoi{10.1093/mnras/stw426}

\bibitem[{Ye {et~al.}(2019)Ye, Kremer, Chatterjee, Rodriguez, \&
  Rasio}]{Ye_2019}
Ye, C.~S., Kremer, K., Chatterjee, S., Rodriguez, C.~L., \& Rasio, F.~A. 2019,
  \apj, 877, 122, \dodoi{10.3847/1538-4357/ab1b21}

\bibitem[{{Ye} {et~al.}(2021){Ye}, {Kremer}, {Rodriguez}, {Rui}, {Weatherford},
  {Chatterjee}, {Fragione}, \& {Rasio}}]{Ye_2021}
{Ye}, C.~S., {Kremer}, K., {Rodriguez}, C.~L., {et~al.} 2021, arXiv e-prints,
  arXiv:2110.05495.
\newblock \doarXiv{2110.05495}

\bibitem[{Zaris {et~al.}(2020)Zaris, Veske, Samsing, M{\'{a}}rka, Bartos, \&
  M{\'{a}}rka}]{Zaris_2020}
Zaris, J., Veske, D., Samsing, J., {et~al.} 2020, \apj, 894, L9,
  \dodoi{10.3847/2041-8213/ab89a3}

\bibitem[{Zevin {et~al.}(2019)Zevin, Samsing, Rodriguez, Haster, \&
  Ramirez-Ruiz}]{Zevin_2019}
Zevin, M., Samsing, J., Rodriguez, C., Haster, C.-J., \& Ramirez-Ruiz, E. 2019,
  \apj, 871, 91, \dodoi{10.3847/1538-4357/aaf6ec}

\bibitem[{Ziosi {et~al.}(2014)Ziosi, Mapelli, Branchesi, \&
  Tormen}]{Ziosi_2014}
Ziosi, B.~M., Mapelli, M., Branchesi, M., \& Tormen, G. 2014, \mnras, 441,
  3703, \dodoi{10.1093/mnras/stu824}

\end{thebibliography}
\end{document}